\documentclass[a4paper,10pt]{article}
\pdfoutput=1
\usepackage[utf8]{inputenc}
\usepackage{amsmath, graphicx,color}
\usepackage{lineno}

\usepackage[top=3cm, bottom=3cm, left=3cm, right=3cm]{geometry}
\definecolor{webblue}{rgb}{0, 0, 0.5} 
\usepackage[pdfpagelabels,  
plainpages=true,
bookmarks=true,
bookmarksnumbered=true,
hypertexnames=true,
breaklinks=true,%
linkcolor=webblue,]{hyperref}
\hypersetup{
colorlinks  = true,
linkcolor = webblue,
urlcolor = webblue,
citecolor = webblue,
pdfauthor   = {John F. Rudge},
pdfsubject = {},
pdftitle    = {Textural equilibrium melt geometries around 
tetrakaidecahedral 
grains},
pdfkeywords = {},
}

\renewcommand{\d}{\mathrm{d}}

\newcommand {\pdd}[2]{\frac{\partial #1}{\partial #2}}

\begin{document}
\title{Textural equilibrium melt geometries around tetrakaidecahedral 
grains}
\author{John F. Rudge\\
\small{Bullard Laboratories, Department of Earth Sciences, 
University  of 
 Cambridge}}

\maketitle

\begin{abstract}
In textural equilibrium, partially molten materials minimise the total 
surface energy bound up in grain boundaries and grain-melt interfaces. 
Here, numerical calculations of such textural equilibrium geometries are 
presented for a space-filling tessellation of grains with a 
tetrakaidecahedral 
(truncated octahedral) unit cell. Two parameters determine the nature of 
the geometries: the porosity and the dihedral angle. A variety of distinct 
melt topologies occur for different combinations of these two parameters, 
and the boundaries between different topologies has been determined. 
For small dihedral angles, wetting of grain boundaries occurs once the 
porosity has exceeded 11\%. An exhaustive account is given of the main 
properties of the geometries: their energy, pressure, mean curvature, 
contiguity, and areas on cross-sections and faces. Their effective 
permeabilities have been calculated, and demonstrate a transition 
between a quadratic variation with porosity at low porosities to a cubic 
variation at high porosities.
\end{abstract}

\section{Introduction}

The physical properties of partially molten materials depend crucially on 
the 
geometry of melt at the scale of individual grains. Properties 
like permeability or electrical conductivity can be radically 
different depending on whether melt forms a connected network or not. The 
aim of this contribution is to better understand the controls on the 
geometry of melt networks in order to ultimately better understand the 
physical properties of partially molten materials.

Surface energy plays a key role in determining melt 
network geometry. In the absence of external forcing, partially molten 
materials tend 
to a state of textural equilibrium in which the surface energies bound up 
in grain boundaries and grain-melt interfaces are minimised. While in many 
situations a state of textural equilibrium is not achieved (due to 
the action 
of additional mechanical and chemical processes), it provides an important 
reference model for understanding melt geometry.

It is straightforward to write down a mathematical statement of 
textural equilibrium, but solving the resulting equations is much less 
straightforward. Early work \cite{Beere1975, Wray1976, 
Bulau1979, Park1985} provided analytical solutions in some simple special 
cases, or used fairly rough approximations to the geometry. Only in the 
late 1980s, was the 
simplest 3-dimensional problem -- four grains meeting at junction with 
tetrahedral symmetry -- solved fully numerically, by 
von Bargen and Waff \cite{vonBargen1986}, Cheadle \cite{Cheadle1989}, 
and Nye \cite{Nye1989}. This 
simple model with tetrahedral symmetry provides important insights into 
when a melt network is expected to be connected, and also provides 
constraints on the expected permeability \cite{vonBargen1986,Cheadle1989,
Cheadle2004} 
and electrical conductivity \cite{Cheadle1989,Pervukhina2008} of such 
networks. 
However, the tetrahedral-symmetry junctions have an important drawback: 
there is no space-filling solid phase compatible with such junctions. It 
is 
thus difficult to use the results from these 
studies in models which need to describe processes occurring within 
individual solid grains (e.g. as needed in modelling diffusion creep).  

This manuscript provides an exhaustive account of textural equilibrium 
melt geometries around a particular choice of solid grains which do fill 
space. In the absence of melt, these chosen grains take the shape of 
plane-faced tetrakaidecahedrons (truncated octahedrons). The problem has a 
high degree of symmetry, and is only marginally more complex that the 
problem tackled by von Bargen and Waff \cite{vonBargen1986}, 
Cheadle \cite{Cheadle1989}, and 
Nye \cite{Nye1989}. Indeed, this problem has already been tackled in a 
series 
of recent contributions by Ghanbarzadeh et al.
\cite{Ghanbarzadeh2014,Ghanbarzadeh2015,Ghanbarzadeh2015b,Ghanbarzadeh2016,Ghanbarzadeh2017}
, using a novel 
level-set method that allows for greater flexibility in grain 
shape than the approach taken here. However, there are some important 
differences between the results of 
that study and the present study, which will be discussed in detail in 
\autoref{sec:discuss}. The main advantage of the approach taken here over 
that of Ghanbarzadeh et al. \cite{Ghanbarzadeh2014} is that this study 
exploits all the 
symmetries of the problem, which makes it easier to resolve the fine 
details of the melt geometries.

This manuscript is organised as follows: First the model is 
described, along with a recap of some well-known mathematical results 
about textural equilibrium. Accounts are then given of the main properties 
of the melt topologies: energy, pressure, contiguity, and areas on 
contacts 
and cross-sections. Permeability calculations are then discussed, and are 
followed by a more general discussion relating these calculations to 
previous work. Appendices provide more detail on the numerical methods, 
and provide some analytical solutions for special cases.

\section{The model}

\begin{figure}
 \centering
 \includegraphics[width=0.5\columnwidth]{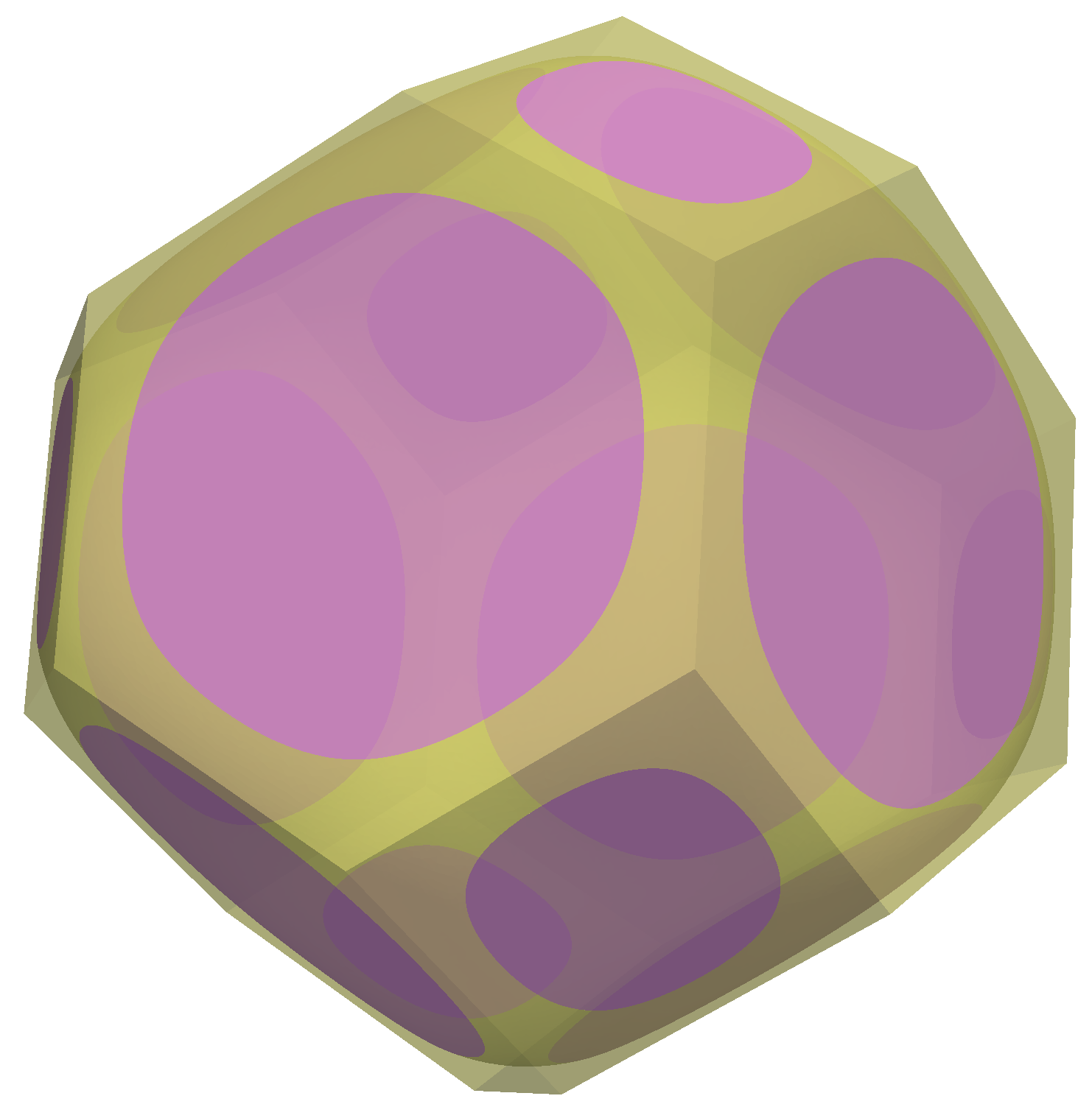}
 \caption{3D rendering of the tetrakaidecahedral unit cell. Grain-grain 
(solid-solid) contacts are shown in pink. Region where melt (liquid) is 
present is shown in yellow. Translucency has been used to allow a view 
through the grain.}
 \label{fig:unit_cell}
\end{figure}

The model geometry consists of an infinite tessellation of 
tetrakaidecahedral unit cells as depicted in \autoref{fig:unit_cell}. In 
the absence of melt, a single grain occupies the whole of a unit cell, and 
the grain takes the shape of a plane-faced tetrakaidecahedron (truncated 
octahedron). The faces of the tetrakaidecahedron form the grain boundaries 
(solid-solid contacts). When melt is present it is assumed that 
grain-boundaries continue to lie along the faces of the tetrakaidecahedral 
cell, although the area and shape of these contacts is allowed to vary. 
Textural equilibrium involves the minimisation of surface energy $E$ 
\cite{Beere1975,Park1985},
\begin{equation}
 E =\frac{1}{2} \gamma_\text{ss} A_\text{ss}  + \gamma_\text{sl} 
A_\text{sl}
\label{eq:energy}
\end{equation}
where $\gamma_\text{ss}$ and $\gamma_\text{sl}$ are the surface energies 
per unit of area of grain-grain (solid-solid) contacts and grain-melt 
(solid-liquid) contacts respectively. $A_\text{ss}$ and $A_\text{sl}$ are 
the corresponding areas of solid-solid and solid-liquid contacts per unit 
cell. In this 
work it will be assumed that $\gamma_\text{ss}$ and $\gamma_\text{sl}$ are 
isotropic and constant.  The factor of $1/2$ in \eqref{eq:energy} arises 
from the fact that the grain boundaries are on the faces of the unit cell, 
so per-unit cell each solid-solid contact only counts for half in the 
total surface energy \cite{Park1985}. 

\begin{figure}
 \centering
 \includegraphics[width=\columnwidth]{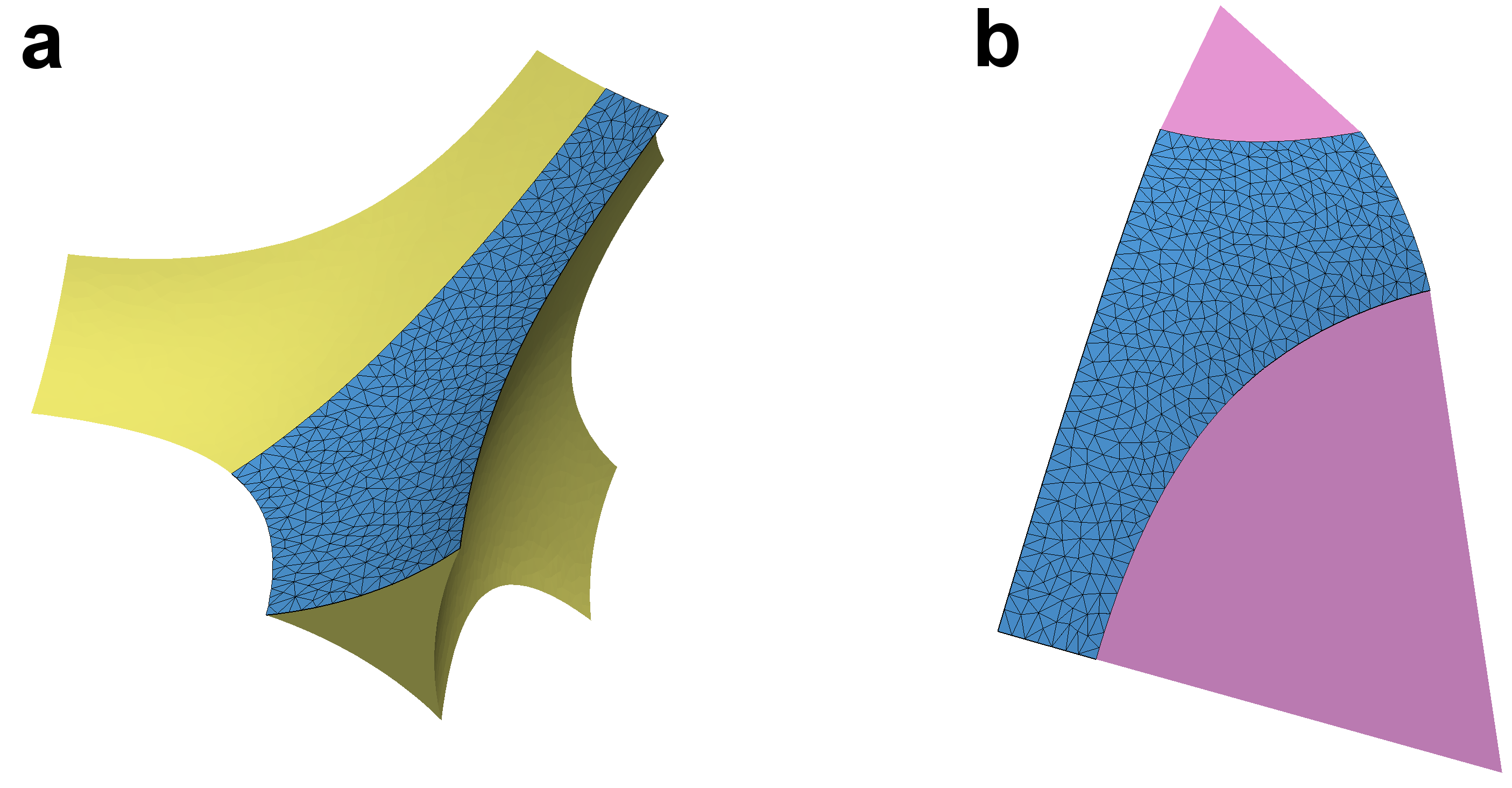}
 \caption{Images showing the fundamental computational domain for a 
porosity $\phi=0.03$ and dihedral angle $\theta = 30^\circ$. a) Melt 
quadruple junction showing solid-liquid interface with fundamental 
computational domain in blue with finite element 
triangulation in black. The rest of the melt junction can be obtained from 
the 
fundamental domain by applying the 8 elements of the point group 
$D_{2d}$ (tennis-ball symmetry). b) Same fundamental computational domain 
in blue, but shown in pink are the solid-solid contacts (grain boundaries) 
associated with the fundamental domain. The full grain can be produced by 
application of the 48 elements of the point group $O_h$. }
 \label{fig:comp_domain}
\end{figure}

Minimisation of $E$ is performed subject to constraints, which reflect 
assumptions about the geometry. Here, grain centres are assumed to reside 
on a body-centred cubic (bcc) lattice, all grain boundaries are assumed to 
be 
planar, and all grains are identical. The unit cell shown in 
\autoref{fig:unit_cell} is the Wigner-Seitz cell of the bcc lattice of 
grain centres. As a consequence of the cubic 
symmetry, the basic computational domain can be 
reduced to a single region that is 1/48th of a grain or 1/8th of a 
quadruple junction (\autoref{fig:comp_domain}). 

Grain centres are fixed in these calculations, and thus there is a 
constant 
distance $d$ between opposing square faces in the unit tetrakaidecahedral 
cell. The volume of the unit cell is $V_\text{cell} = d^3 / 2$ and it has 
area $A_\text{cell} = \frac{3}{4} \left( 1 + 2 \sqrt{3} \right) d^2$. Each 
edge has length $a = d/\left(2 \sqrt{2} \right)$. The volume fraction of 
melt is 
prescribed, by introducing a Lagrange multiplier $\lambda$ and minimising 
the 
functional $J$,
\begin{equation}
 J = E + \lambda \left( V_\text{l} - V_\text{l}^* \right) \label{eq:J}
\end{equation}
where $V_\text{l}$ is the volume of liquid in the domain, and 
$V_\text{l}^*$ is the 
target (prescribed) value. 

The variational problem is discretised by representing the surface by a 
finite element mesh of triangles, using \textit{The Surface 
Evolver} software
\cite{Brakke1992,Brakke1996,Brakke1997}. Numerical optimization is used 
to find the unknown mesh node co-ordinates that minimise the surface 
energy subject to the given constraints. Further details on the numerical 
methods can be found in Appendix \ref{sec:numerics}. 

At a minimum, $\delta J=0$, and from variational calculus it follows that 
at the minimum, $\lambda = \lambda^*$
where
\begin{equation}
 \lambda^* = 2 \gamma_\text{sl} H. \label{eq:Young-Laplace}
\end{equation}
$H$ is the mean curvature of the solid-liquid surface, defined by
\begin{equation}
 H \equiv \tfrac{1}{2} \left( {1}/{R_1} + {1}/{R_2} \right) \equiv 
\nabla_s \cdot \mathbf{n}
\end{equation}
where $R_1$ and $R_2$ are the principal radii of curvatures, and 
$\mathbf{n}$ is the normal vector (chosen here to point outward from solid 
/ inward to liquid). $\nabla_s \cdot$ represents the surface divergence 
operator. Thus in textural equilibrium, the mean curvature of the 
solid-liquid surface is a constant. Moreover, the Lagrange multiplier 
enforcing the volume constraint has a physical interpretation: it 
represents the pressure difference between solid and liquid, $\lambda^* = 
\Delta P = P_\text{s} - P_\text{l}$, and \eqref{eq:Young-Laplace} is the 
Young-Laplace 
equation relating pressure differences to mean curvature.

\begin{figure}
 \centering
 \includegraphics[width=0.75\columnwidth]{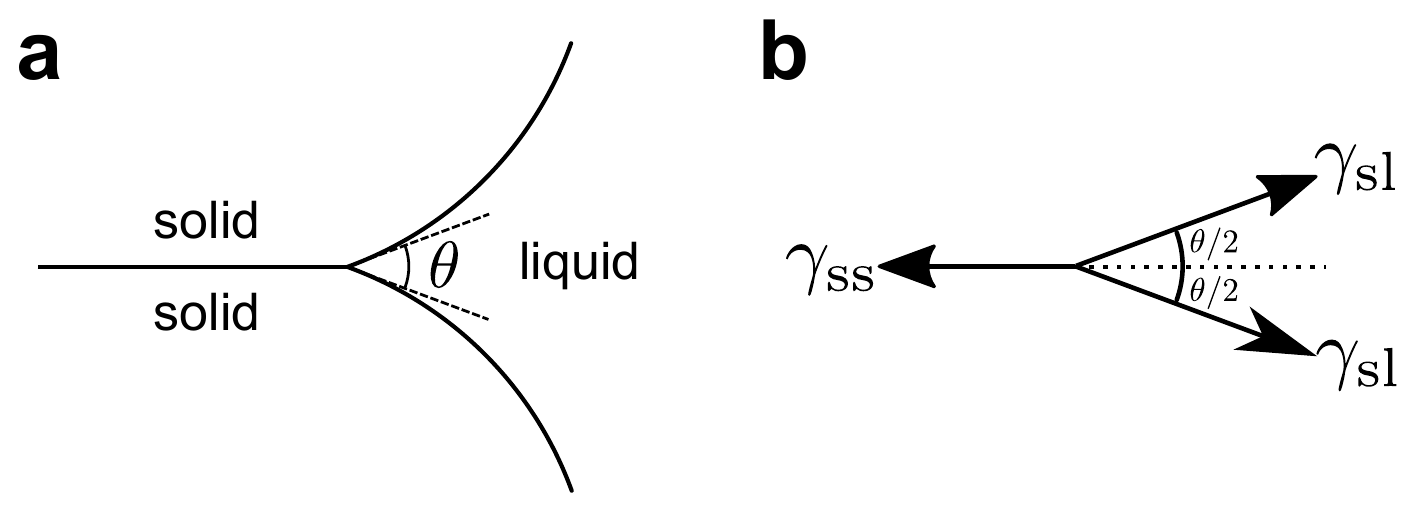}
 \caption{a) Illustration of the dihedral angle $\theta$ as the angle at which two solid-liquid surfaces meet a solid-solid surface. b) The corresponding force balance at the triple line.}
 \label{fig:di_angle}
\end{figure}

Also from $\delta J=0$ it follows that the following force balance holds 
along triple lines where three surfaces meet \cite{Smith1948}, 
\begin{equation}
 \sum_{i=1}^3 \gamma_i \boldsymbol{\nu}_i = \mathbf{0}
\end{equation}
where $\boldsymbol{\nu}_i$ is the co-normal to each surface at the triple 
line, and $\gamma_i$ the corresponding surface energy per unit area. This 
reduces to the familiar expression for the dihedral angle $\theta$ at 
which 
two solid-liquid surfaces meet a solid-solid surface  (\autoref{fig:di_angle}),
\begin{equation}
 \cos \frac{\theta}{2} = \frac{\gamma_\text{ss}}{2 \gamma_\text{sl}}.
\end{equation}

It should be noted that the approach taken here, of discretising the 
variational 
problem (minimise $J$ in \eqref{eq:J}), differs from the approach taken in 
the 
classic 
studies by von Bargen and Waff \cite{vonBargen1986}, 
Cheadle \cite{Cheadle1989}, and Nye 
\cite{Nye1989}, and 
also the more recent work by Ghanbarzadeh et al. \cite{Ghanbarzadeh2014}. 
The starting point 
for 
all of these studies is the statement that the solid-liquid surfaces are 
of 
constant mean curvature and meet the solid-solid surfaces at the dihedral 
angle. These studies are all based on a numerical discretisation of the 
mean curvature of the solid-liquid surfaces, whereas here in discretising 
$J$ only areas and volumes are discretised. The approach 
taken 
here, working from the variational problem, is essentially identical to 
the 
approach pioneered by Beere \cite{Beere1975}, although here a more 
refined 
discretisation is used. 

By scaling, the variational problem can be reduced to being a function of 
just two dimensionless parameters: the porosity $\phi$ (the volume 
fraction of 
melt), and the dihedral angle $\theta$. For example, if $E$ represents the 
surface energy contained in one tetrakaidecahedral cell, then a scaled 
energy 
can be defined by $E^\prime = E /(\gamma_\text{sl} A_\text{cell})$ where 
$A_\text{cell}$ is the area of the bounding tetrakaidecahedral cell 
($A_\text{cell} = \frac{3}{4} \left( 1 + 2 \sqrt{3} \right) d^2$). All 
results 
in this manuscript are presented using scaled variables as functions of 
$\phi$ 
and $\theta$.

\section{Melt topologies}

\begin{figure}
 \centering
 \includegraphics[width=\columnwidth]{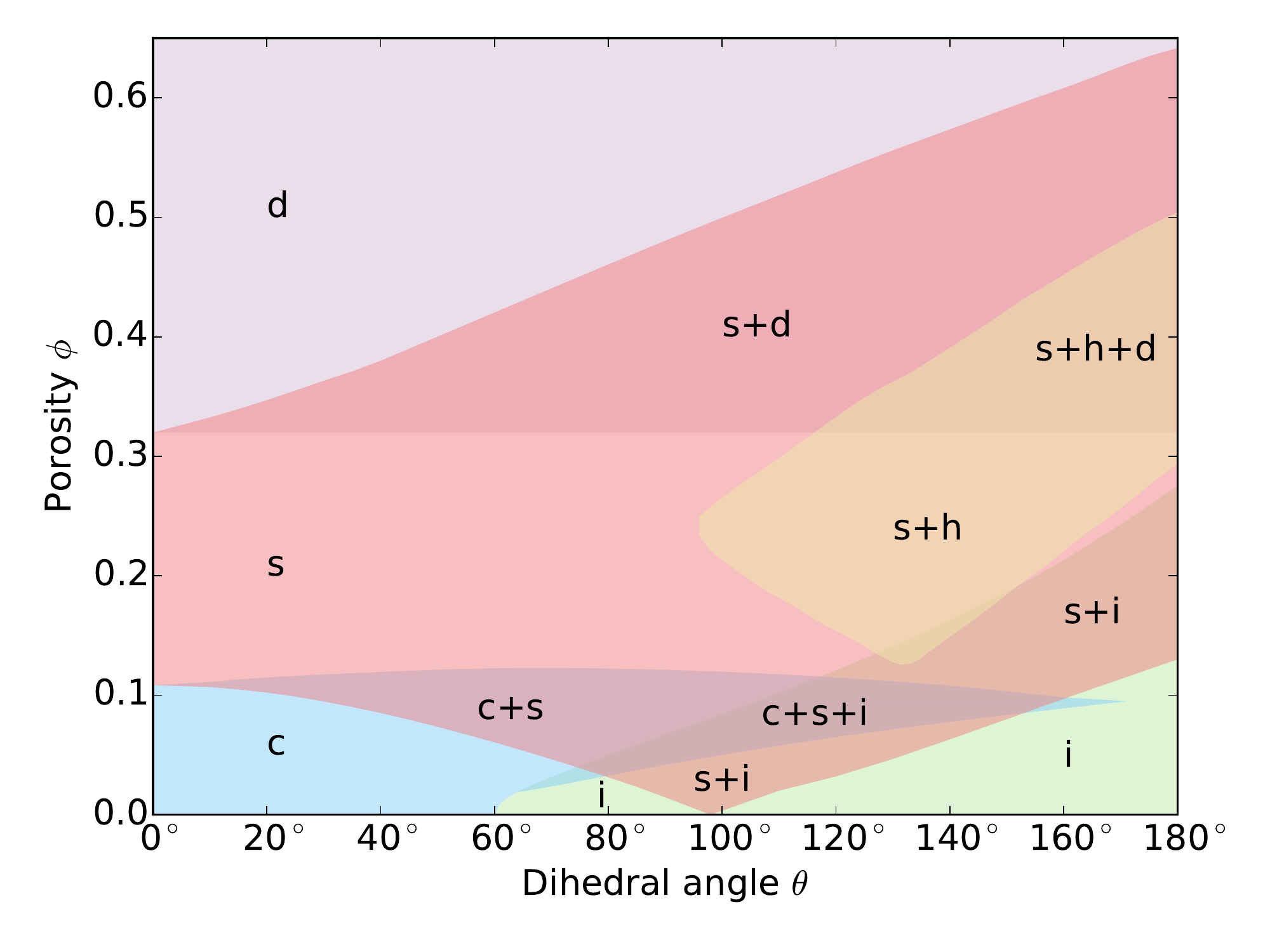}
 \caption{Regime diagram showing regions where different melt topologies 
have 
been found to exist. ``c'' denotes melt connected along grain edges 
(\autoref{fig:connected}). ``i'' denotes melt in isolated pockets at the 
grain 
corners (\autoref{fig:isolated}). ``s'' denotes melt which wets the square 
faces of the grains, but not the hexagonal faces 
(\autoref{fig:wet_square}). 
``h'' denotes melt which wets the hexagonal faces, but not the square 
faces 
(\autoref{fig:wet_hex}). ``d'' denotes disaggregation, where grains are 
isolated spheres surrounded by melt. Pluses indicate regions where 
multiple topologies are found.}
 \label{fig:regime}
\end{figure}

One of the key features of textural equilibrium is that different kinds of 
melt 
topology are possible for different values of the dihedral angle and at 
different porosities \cite{Wray1976, Bulau1979}. For example, there is 
the well-known result that at 
small 
porosities, the melt network is only connected along the grain edges for 
dihedral angles less than $60^\circ$. The problem tackled here allows for 
a 
rich variety of melt topologies as a function of porosity and 
dihedral angle, which are summarised in the regime diagram of 
\autoref{fig:regime}, and discussed below. For some combinations of 
parameters it is possible to find multiple topologies: each is a local 
minimum of the energy functional, but not necessarily a global minimum. 

\begin{figure}
 \centering
 
\includegraphics[width=\columnwidth]{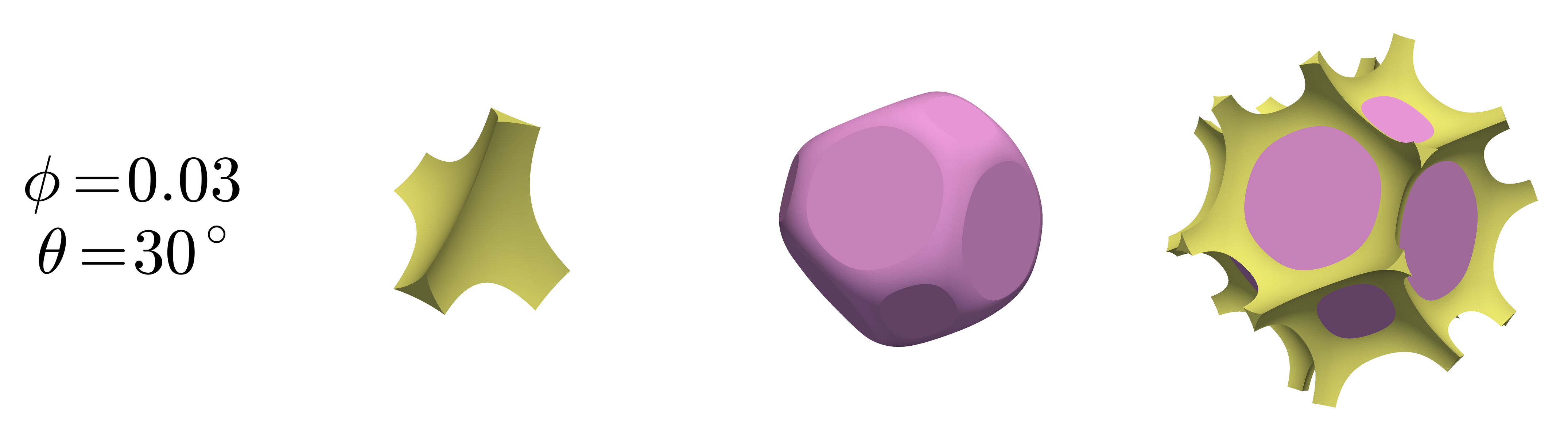}
\\
 
\includegraphics[width=\columnwidth]{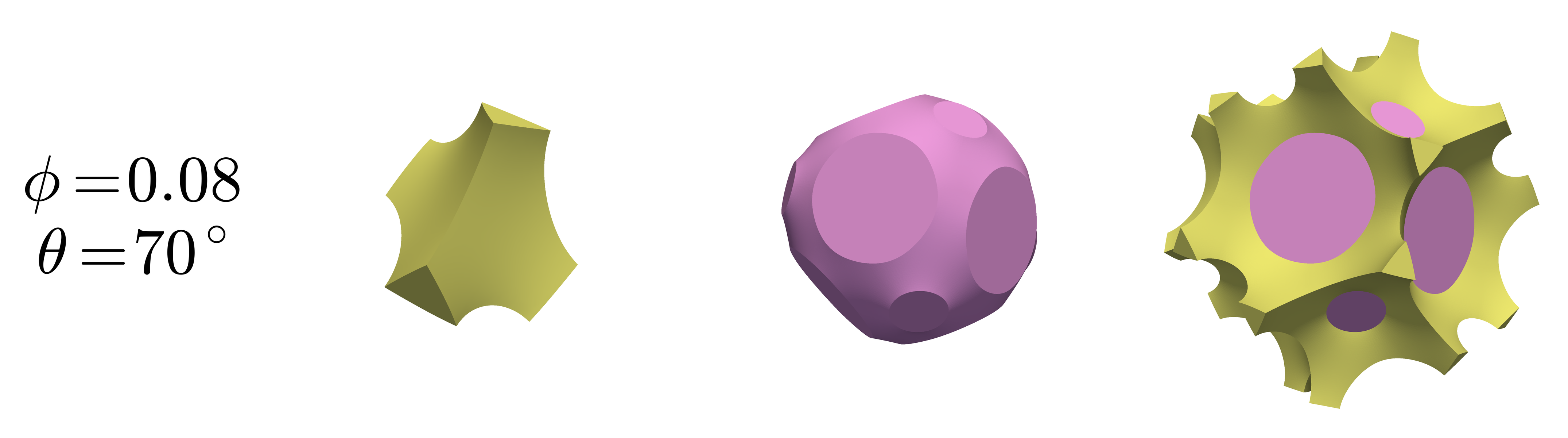}
\\
\includegraphics[width=\columnwidth]{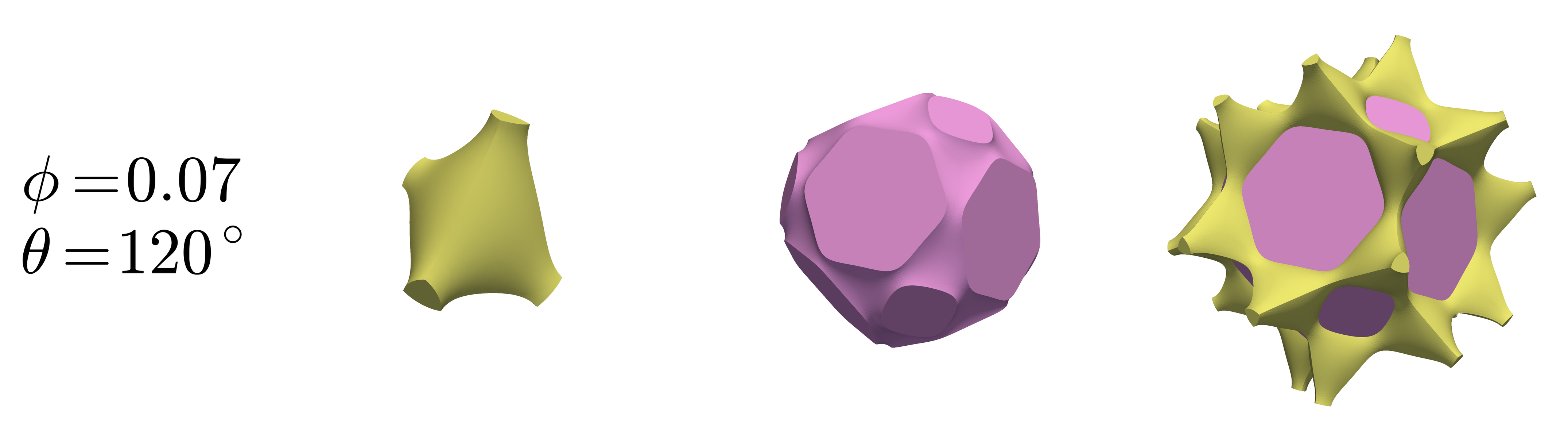}
\\
 \caption{Examples of melt topology ``c'', connected along grain edges. In 
yellow on left is a view of the quadruple junction. In pink in middle is a 
view 
of an individual grain. On right is a view of a individual grain + melt. 
Corresponding labels give porosity $\phi$ and dihedral angle $\theta$. The 
first two rows have positive mean curvature, the final row has negative 
mean curvature.}
 \label{fig:connected}
\end{figure}

\begin{figure}
 \centering
 \includegraphics[width=\columnwidth]{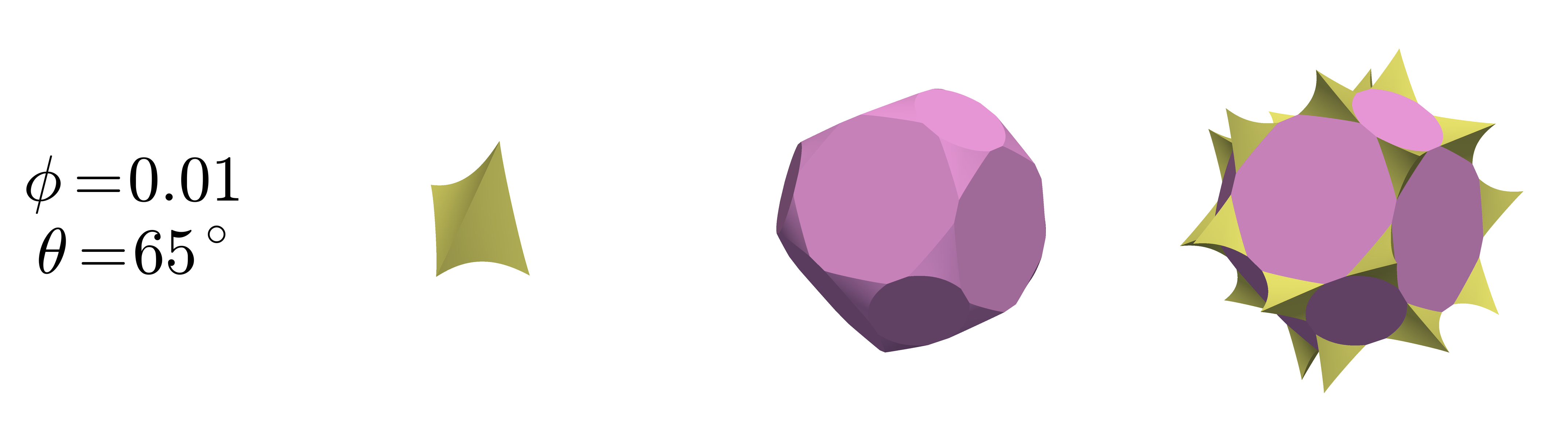}
\includegraphics[width=\columnwidth]{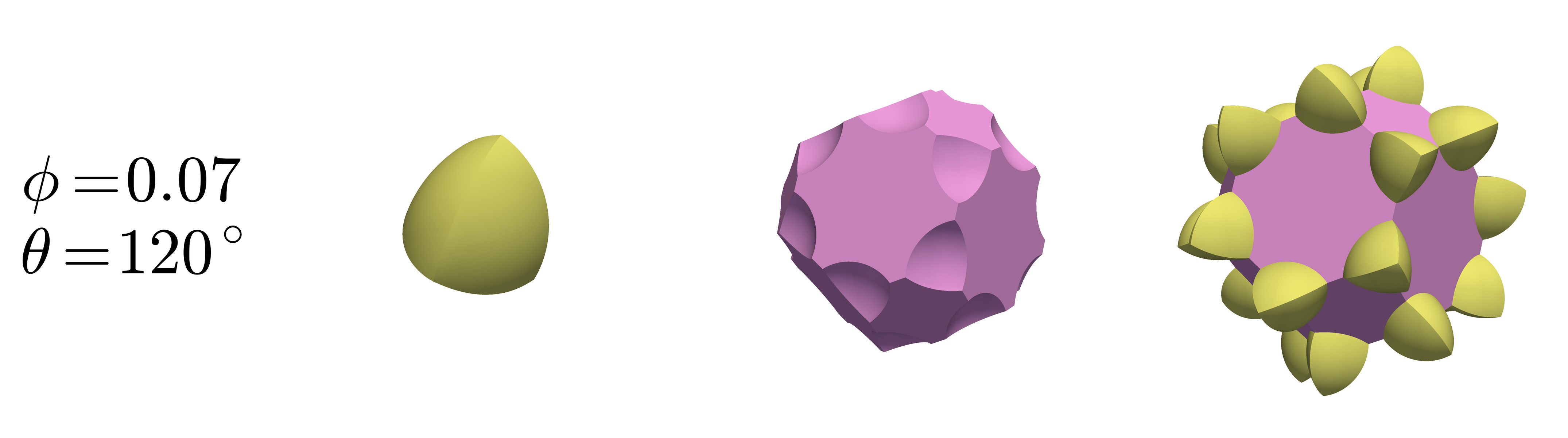}
\\
 \caption{Examples of melt topology ``i'', where melt is isolated at the 
grain corners. Note that $\theta=65^\circ$ has positive mean curvature 
whereas $\theta=120^\circ$ has negative mean curvature. Note that the 
bottom row has the same combination of porosity and dihedral angle as the 
bottom row of \autoref{fig:connected}.}
 \label{fig:isolated}
\end{figure}
The first type of topology, marked ``c'' in \autoref{fig:regime}, has melt 
forming a connected network along the grain edges. Examples of such a 
topology 
are shown in \autoref{fig:connected}, and are broadly similar to those 
depicted 
by von Bargen and Waff \cite{vonBargen1986}, Cheadle \cite{Cheadle1989} 
and Nye \cite{Nye1989}, except 
that 
the melt junction does not have tetrahedral symmetry. The second type of 
topology, ``i'' (\autoref{fig:isolated}), which consists of melt isolated 
at 
grain corners is also broadly similar to previous calculations which 
assumed 
tetrahedral symmetry, for which analytical expressions are available 
\cite{Wray1976}. At very large porosities the grain-grain contacts 
disappear 
and grains are completely surrounded by melt. This is marked as ``d'' for 
disaggregated in \autoref{fig:regime}, where the minimum energy 
configuration 
simply consists of spherical grains. ``d'' topologies exist for all 
porosities 
greater than $1- \pi \sqrt{3}/8 \approx 32\%$.

\begin{figure}
 \centering
 \includegraphics[width=\columnwidth]{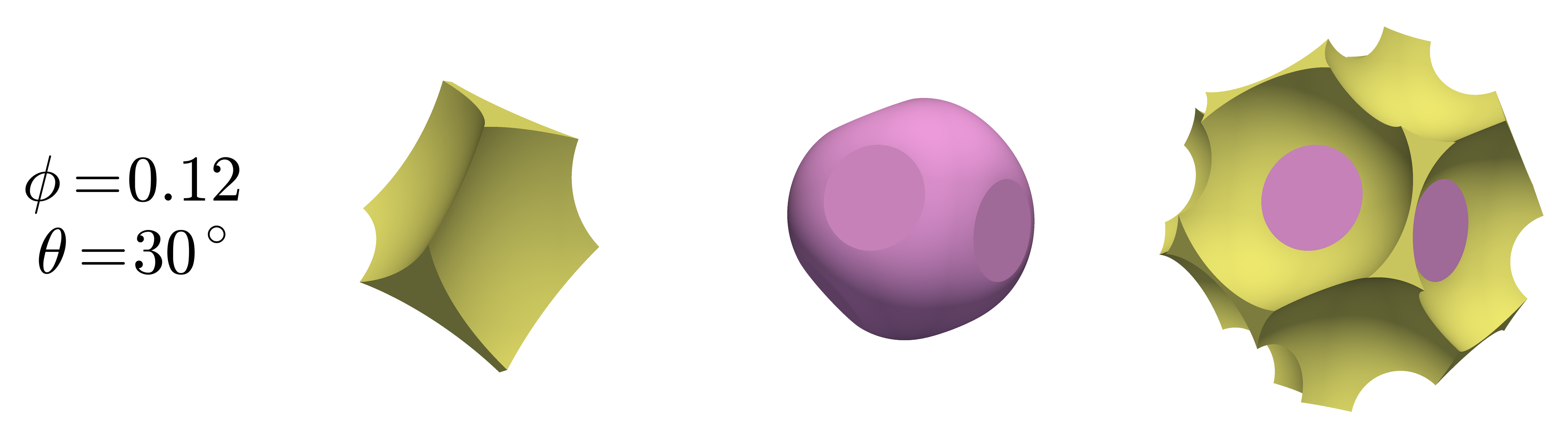}\\
 \includegraphics[width=\columnwidth]{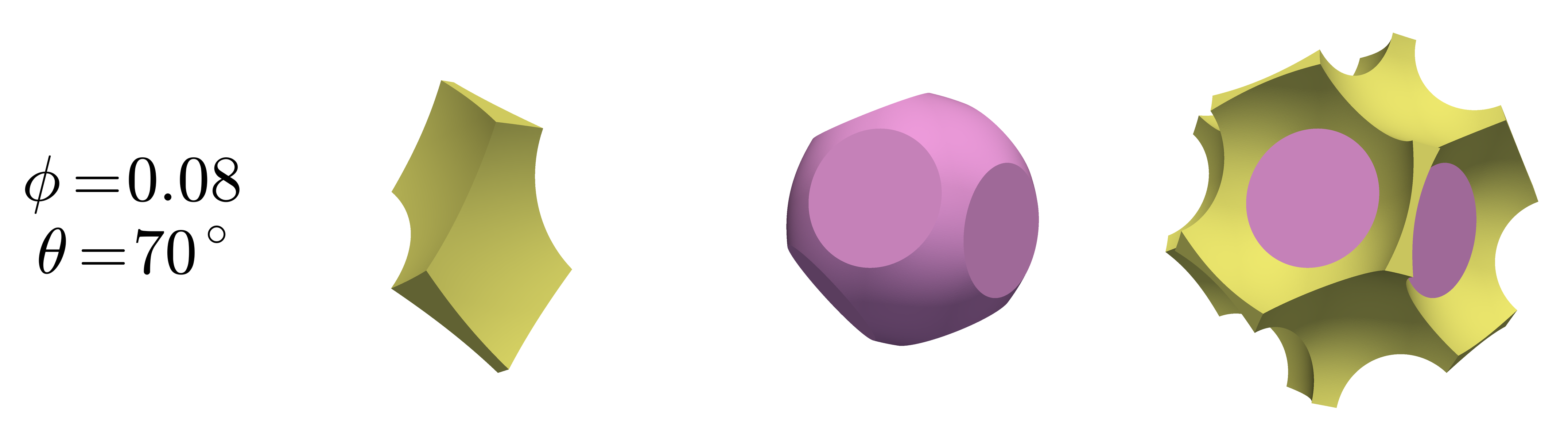}\\
 \includegraphics[width=\columnwidth]{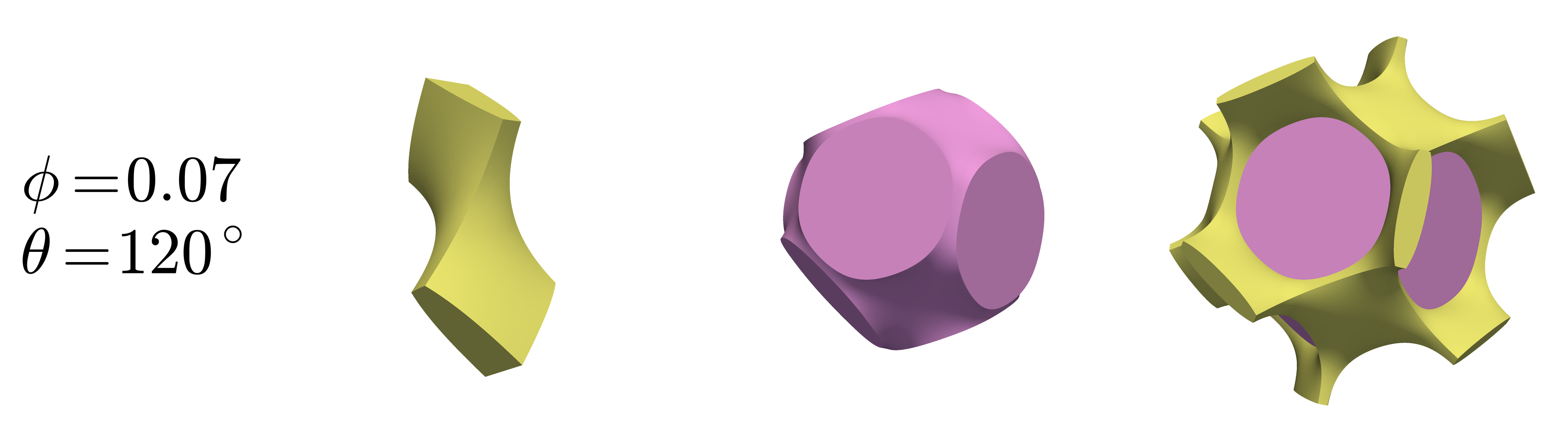}\\
 \caption{Examples of melt topology ``s'', where melt is connected and 
wets 
the square faces. Note that the 
bottom two rows have the same combinations of porosity and dihedral angle 
as the 
bottom two rows of \autoref{fig:connected}.}
 \label{fig:wet_square}
\end{figure}

\begin{figure}
 \centering
 \includegraphics[width=\columnwidth]{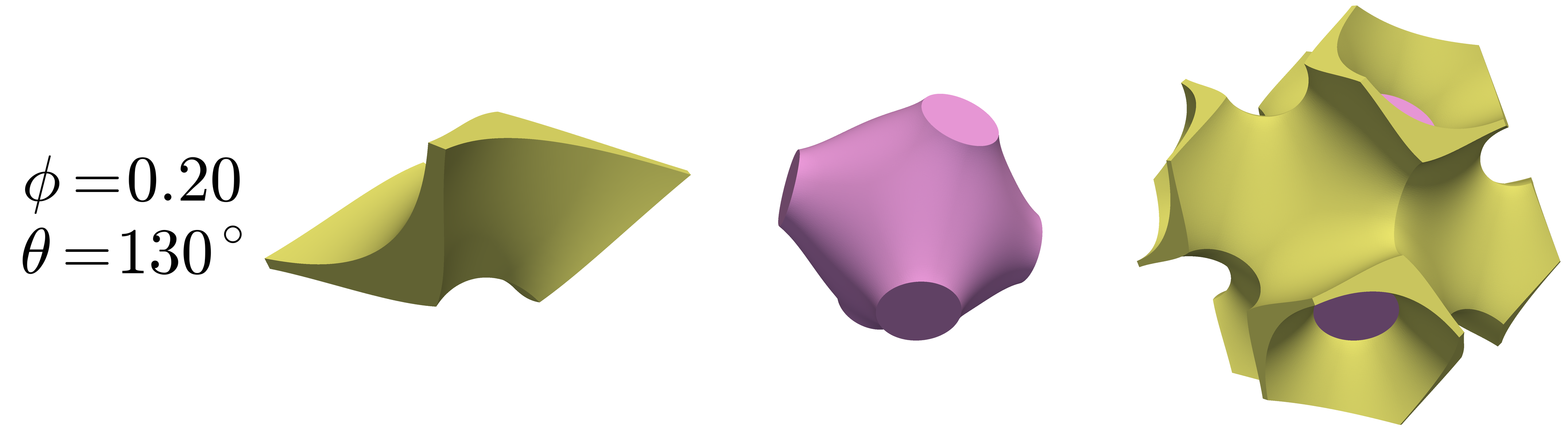}\\
 \caption{An example of melt topology ``h'', where melt is connected and 
wets 
the hexagonal faces.}
 \label{fig:wet_hex}
\end{figure}

The problem considered here admits additional melt topologies because of 
the 
lower degree of symmetry of the quadruple junction. These are depicted 
in Figures \ref{fig:wet_square} and \ref{fig:wet_hex}. The first of these 
represents the case ``s'' where the square faces become wetted, but the 
hexagonal faces do not. The second represents the case ``h'' where the 
hexagonal faces are wetted but the square faces are not.

For low porosities, the dihedral angles for which topologies ``c'' and 
``i'' 
are found are essentially the same as found for the problems with 
tetrahedral 
symmetry: isolated solutions exist only for dihedral angles greater than 
$60^\circ$. For dihedral angles greater than this, there is a region of 
overlap 
between the ``c'' and ``i'' topologies as porosity increases (the region 
between the ``pinch-off'' and ``wetting'' boundaries as described by 
von Bargen and Waff \cite{vonBargen1986}). 

Perhaps the most important new transition in the present study is the 
transition from ``c'' to ``s'' topologies as porosity increases (i.e. the 
preferential wetting of the smaller square faces). For low dihedral 
angles, this 
occurs close to $\phi=0.11$, but for larger dihedral angles there is an 
overlap 
where both ``c'' and ``s'' topologies exist. The wetting of square faces 
at around $\phi=0.11$ is also seen 
in calculations of wet Kelvin foams 
\cite{Kraynik2003,Daxner2006,Koehler2000,Murtagh2015,Phelan1995,
Weaire2003}. Wet Kelvin foams are essentially a limiting case of the 
problem considered here: the case where the dihedral angle approaches 
$0^\circ$. The network of melt considered here is termed a Plateau border 
network in the foam literature.  The only difference is that a wet Kelvin 
foam allows for some curvature along what are the grain-grain contacts in 
the present model, but such curvature is extremely slight, and unlikely to 
significantly affect where the transition to wetted square faces occurs.

One curious feature of \autoref{fig:regime} is that for dihedral angles 
just below $100^\circ$ the ``s'' solutions exist down to very small 
porosities. If such a solution were realisable, there would be the 
potential for percolation of melt at small porosities even for some 
dihedral angles greater than $60^\circ$. However, as will be seen in the 
next section, such ``s'' topologies are higher energy than the ``i'' 
topologies and are thus less likely to be realised. 

It should be noted that \autoref{fig:regime} does not provide a map of all 
possible melt topologies, only a key subset that have chosen to be 
investigated. Firstly, the topologies investigated are only those 
consistent with the symmetries imposed. Topologies with lower degrees of 
symmetry are possible e.g. isolated topologies with melt at some quadruple 
junctions but not others. Moreover, there are additional isolated 
topologies possible that are consistent with the imposed symmetry but that 
have not been calculated here. For example, melt can also be isolated in 
the middle of the grain edges, or at the centres of the grain faces. One 
can also have some combination of isolated melt in all three places -- 
edges, faces, and vertices of the unit cell. However, at the onset of 
melting it is expected that melt first forms at the quadruple junctions, 
so it is the case of melt at these junctions that is typically of most 
interest.

\section{Energy and pressure}

\begin{figure}
 \centering
 \includegraphics[width=\columnwidth]{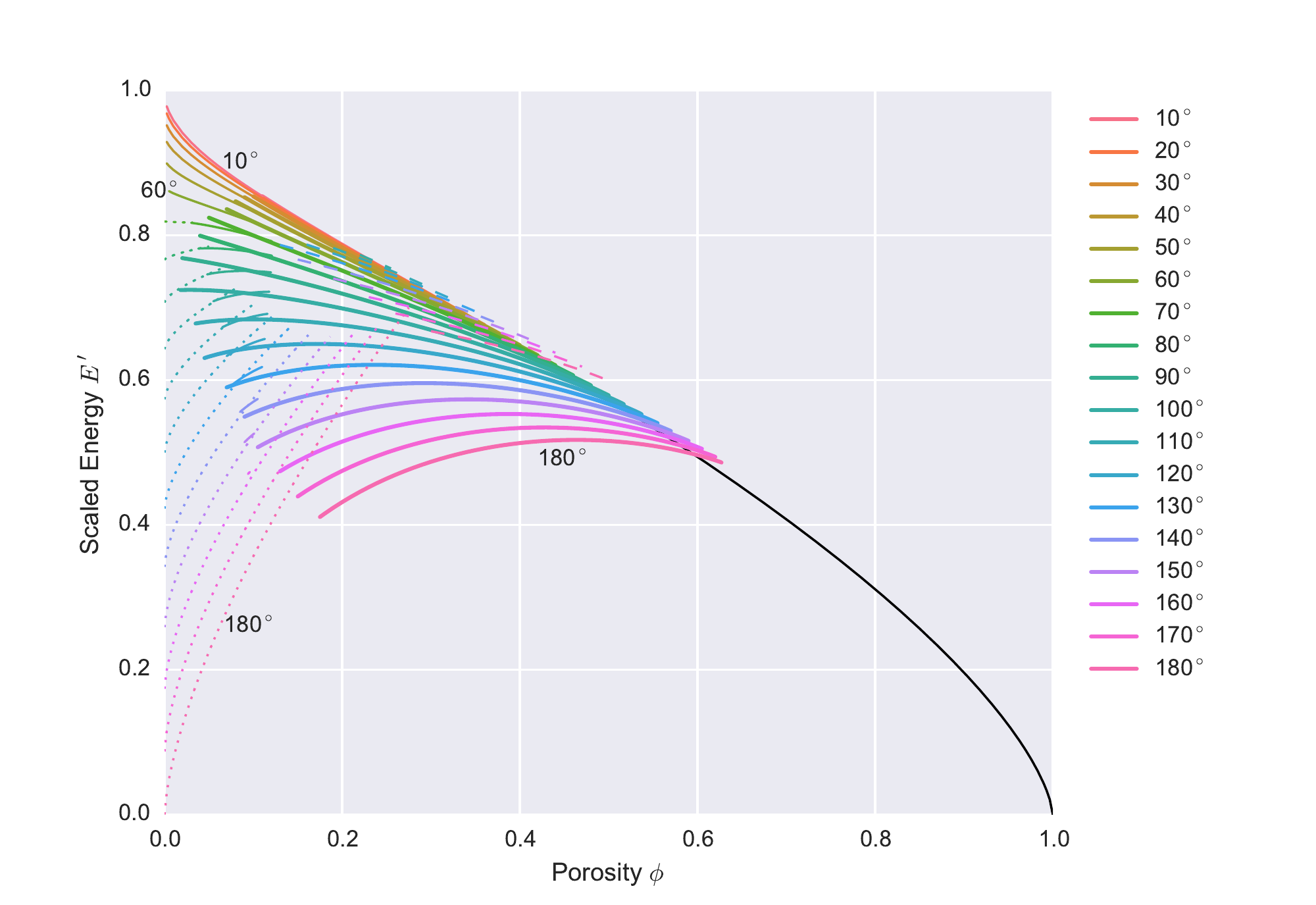}
 \caption{Scaled (dimensionless) energy, $E^\prime = E /(\gamma_\text{sl} 
A_\text{cell})$, plotted as a function of porosity $\phi$ for different 
values of the dihedral angle $\theta$ (legend on right). In this plot, and 
subsequent plots, thin solid lines represent connected topology ``c'', 
thick solid lines square-wetted topology ``s'', dotted lines isolated 
topology ``i'', and dashed lines hex-wetted topology ``h''. The black 
solid line represents disaggregated topology ``d'' (here independent of 
dihedral angle).}
 \label{fig:energy}
\end{figure}

\autoref{fig:energy} plots a scaled surface energy for each of the melt 
topologies as a function of porosity and dihedral angle. For those regions 
of $\phi-\theta$ space that admit multiple topologies, 
\autoref{fig:energy} can be used to identify which topology has the 
overall 
lowest energy (and hence more likely to be found). For example, the ``h'' 
topologies (which wet the hexagonal faces) always have higher energy than 
the ``s'' topologies (which wet the square faces). This what one might 
intuitively expect -- it is more likely for smaller faces to become wetted 
than larger faces.

The situation is more complicated in the region of overlap between ``c'' 
and ``s''. For small dihedral angles, the ``c'' topologies have lower 
energy than the ``s'' topologies for most of the region of overlap, until 
close to the boundary where only ``s'' exists. For larger dihedral angles 
the ``s'' topologies are lower energy for most of the range of overlap. 
Similarly, in the region of overlap between ``s'' and ``i'', for lower 
dihedral angles the ``i'' topologies are lower energy for much the region 
of overlap, whereas at higher dihedral angles the ``s'' topologies are 
lower energy for the whole region of overlap.

\begin{figure}
 \centering
 \includegraphics[width=\columnwidth]{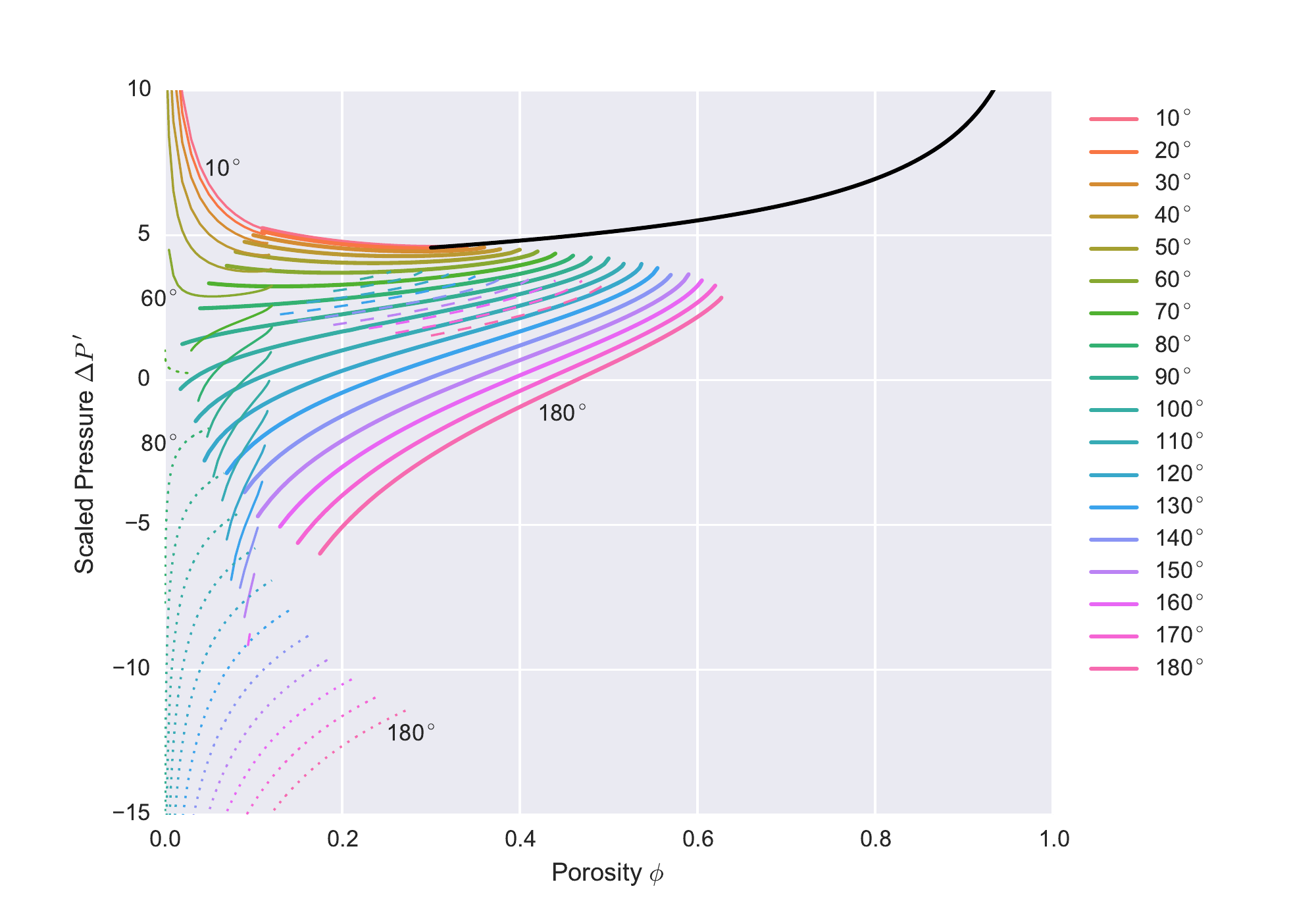}
 \caption{Scaled pressure (or scaled mean curvature), $\Delta P^\prime = 
\Delta P d/ \gamma_\text{sl} = 2 H d$ as a function of porosity.}
 \label{fig:pressure}
\end{figure}

\autoref{fig:pressure} plots a scaled version of the pressure difference 
$\Delta P$ between solid and liquid for the various melt topologies. Owing 
to the Laplace-Young equation \eqref{eq:Young-Laplace}, this is also a 
plot 
of a scaled mean curvature. This pressure difference is calculated during 
the 
solution of the variational problem as $\Delta P = \lambda^*$ is the 
Lagrange multiplier that enforces the fixed volume constraint. The curves 
in \autoref{fig:pressure} are related to the slopes of the curves for 
energy in \autoref{fig:energy}, because of the relationship
\begin{equation}
 \Delta P = - \left. \pdd{E^*}{V^*_\text{l}} \right\vert_{V^*_\text{cell}},
\end{equation}
which is a consequence of the variational problem. Here $E^*$ represents 
the energy at equilibrium, and $V^*_\text{l}$ and $V^*_\text{cell}$ the 
volumes 
of liquid and unit cell. In the scaled (dimensionless) variables used in 
Figures \ref{fig:energy} and \ref{fig:pressure}, this relationship becomes
\begin{equation}
\Delta P^\prime = - \frac{3}{2} \left(1 + 2 \sqrt{3} \right) 
\frac{\partial 
E^\prime}{\partial \phi}.
\end{equation}

\autoref{fig:pressure} shows a number of features that are consistent with 
previous 
studies. For example, for the isolated topologies ``i'', the curvature is 
positive for $\theta < 71^\circ$ and negative for $\theta > 71^\circ$ 
\cite{Wray1976}. Interestingly, topologies ``i'', ``c'', and ``s'' 
have zero mean curvature for some combinations of porosity and dihedral 
angle, and are thus examples of ``minimal surfaces'' (surfaces defined as 
having zero mean curvature). 

The pressure difference between the two phases becomes singular as the 
porosity approaches zero, but the form of singularity varies with the 
dihedral angle. For isolated topologies the porosity dependence is of the 
form $\Delta P \propto \phi^{-1/3}$ (see \eqref{eq:sphere_curvature} in 
\autoref{app:analytical}). For small dihedral angles, the melt geometry 
can 
be closely approximated by tubes along the grain edges, for which $\Delta 
P 
\propto \phi^{-1/2}$ (see \eqref{eq:tube_curvature} in 
\autoref{app:analytical}).

\section{Effective pressure}

\begin{figure}
 \centering
 \includegraphics[width=\columnwidth]{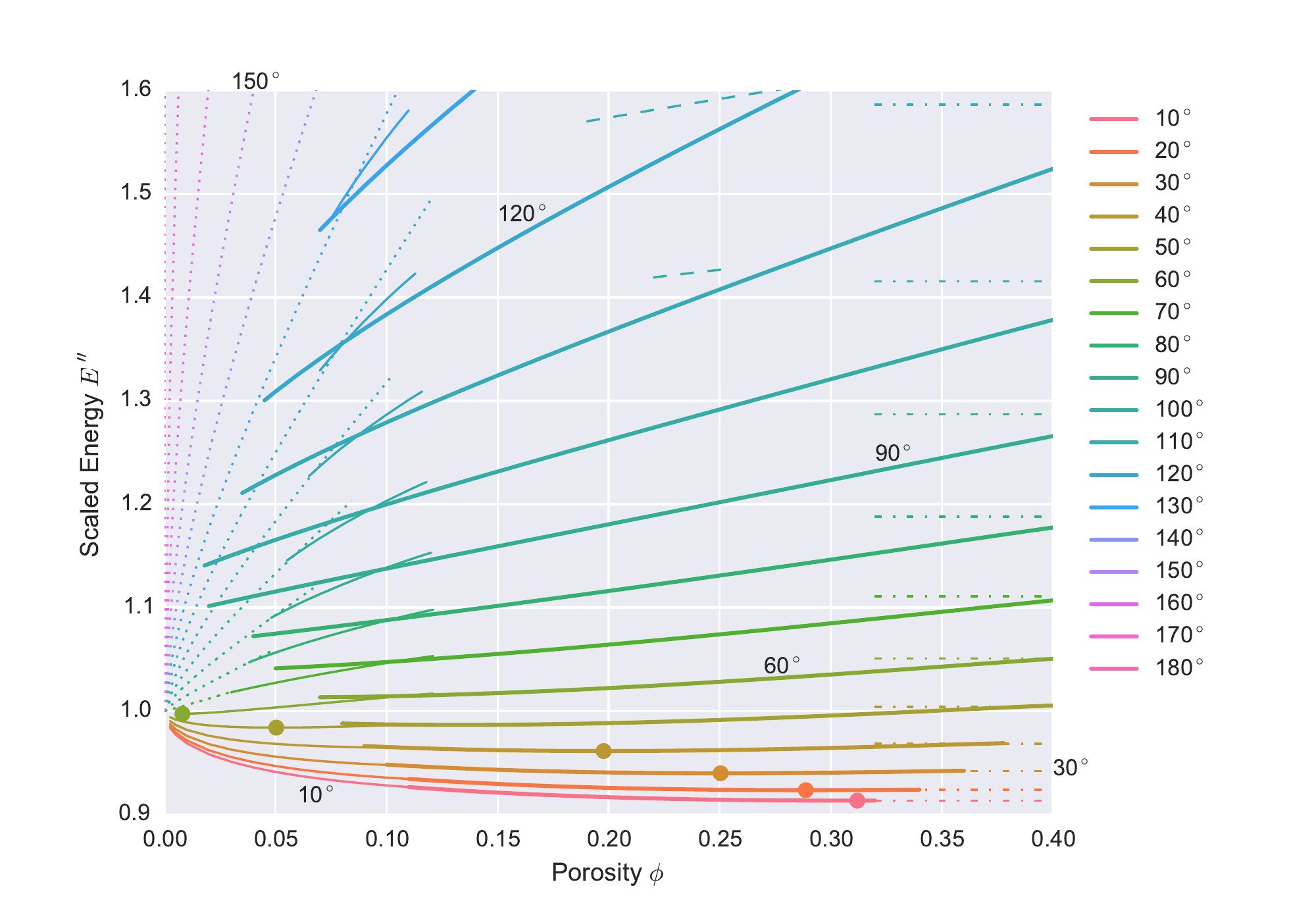}
 \caption{Scaled energy, scaled in the same way as Figure 2 of 
Park and Yoon \cite{Park1985}, $E^{\prime\prime} = E / \left(\tfrac{1}{2} 
\gamma_\text{ss} A_\text{cell0} \right)$. Filled circles show the minimum 
values, which correspond to zero crossings in \autoref{fig:pressure_py}. 
Dash-dotted lines show the energy for the disaggregated topology 
``d'' (which varies with dihedral angle in this scaling, unlike the 
scaling in \autoref{fig:energy}).}
 \label{fig:energy_py}
\end{figure}

\begin{figure}
 \centering
 \includegraphics[width=\columnwidth]{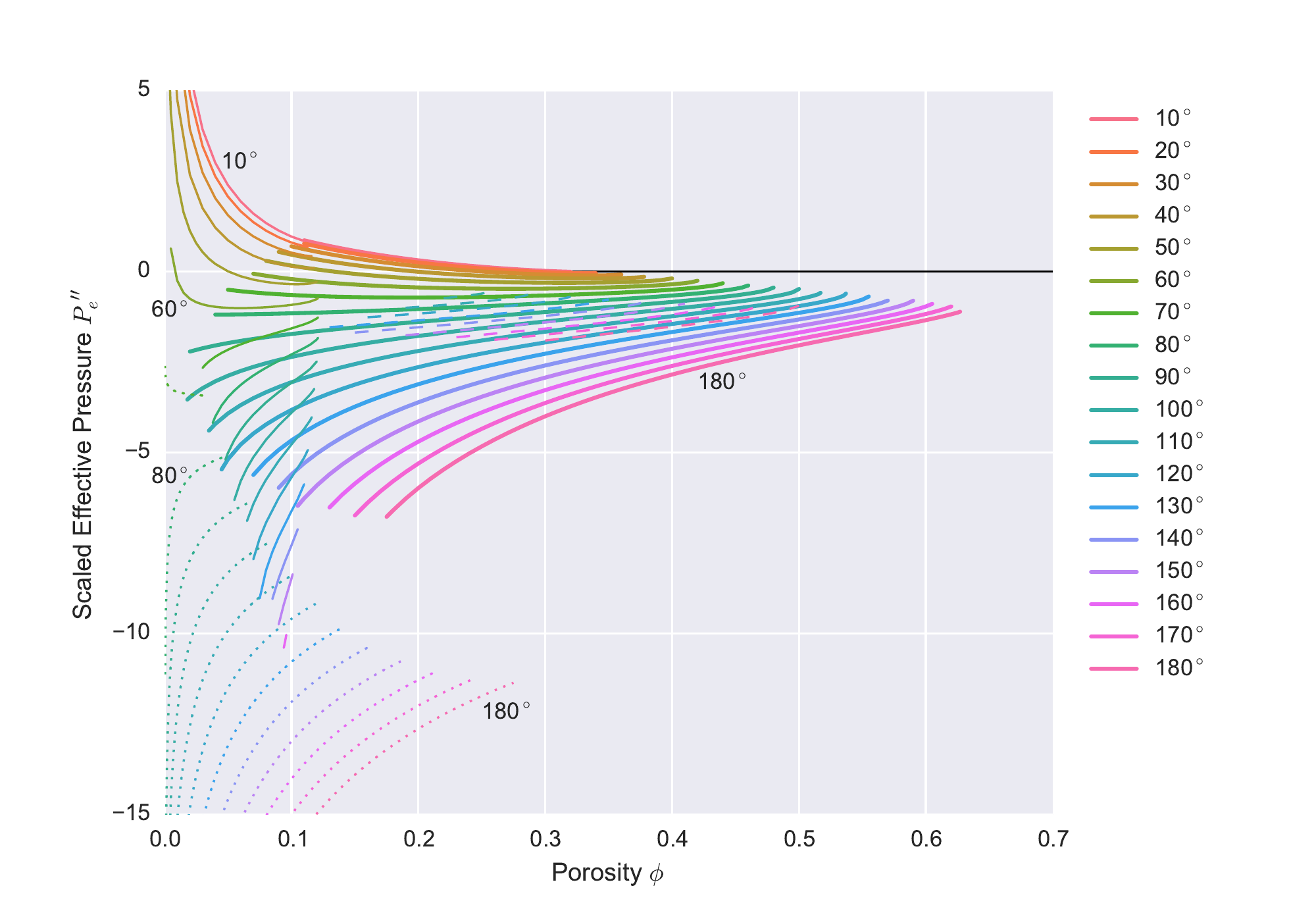}
 \caption{Scaled effective pressure, scaled in the same way as Figure 4 of 
Park and Yoon \cite{Park1985}, $P_e^{\prime \prime} = P_e d_0 / 
\gamma_\text{sl}  
$.}
 \label{fig:pressure_py}
\end{figure}

The pressure plotted in \autoref{fig:pressure} is a measure of the change 
in energy of the system with liquid content, assuming the cell size 
remains 
fixed (and thus as the volume of liquid increases, the volume of solid 
must decrease). As pointed out by 
Park and Yoon \cite{Park1985}, in several physical situations what one 
would like to 
know is the change in energy with liquid content, but keeping the solid 
volume ($V_s^*$) fixed. For example, in problems of compaction or 
sintering 
one may have a partially molten media surrounded by a reservoir of 
liquid, and want to know whether liquid will be drawn in or expelled from 
the partially molten media. Park and Yoon \cite{Park1985} quantify this 
in terms of an 
``effective pressure'', $P_e$, defined by
\begin{equation}
 P_e = - \left. \pdd{E^*}{V^*_l} \right\vert_{V^*_s}.
\end{equation}
Figures \ref{fig:energy_py} and \ref{fig:pressure_py} show scaled energy 
and scaled effective pressure, analagous to Figures 2 and 4 of 
Park and Yoon \cite{Park1985}. They are scaled such that, as porosity 
varies, the solid 
volume remains fixed (rather than the cell size). Scaling is made with 
respect to the length $d_0$,
\begin{equation}
 d_0 = d \left(1- \phi \right)^{1/3},
\end{equation}
which is defined as the distance between grain centres in the absence of 
melt with the same volume of solid grain. The area of the unit cell in the 
absence of melt is given as $A_\text{cell0} = \frac{3}{4} \left( 1 + 2 
\sqrt{3} \right) d_0^2$, and the volume as $V_\text{cell0} = d_0^3 / 2$. 
The plot of effective pressure in \autoref{fig:pressure_py} is related to 
the slope of \autoref{fig:energy_py} by
\begin{equation}
P_e^{\prime\prime} = - \frac{3}{2} \left(1 + 2 \sqrt{3} \right) 
\left(1- \phi \right)^2    \frac{\partial 
E^{\prime \prime}}{\partial \phi} \cos \frac{\theta}{2}.
\end{equation}
If the sign of $P_e$ is positive it indicates there is a driving force 
sucking liquid into the partially molten material. If negative, the 
driving force acts to expel liquid. Zero $P_e$ represents a stable state 
(a 
minimum of the scaled energy plotted in \autoref{fig:energy_py}). The key 
result 
of Figures \ref{fig:energy_py} and  \ref{fig:pressure_py} is that 
topologies for dihedral angles greater than $60^\circ$ are always 
unstable, 
and the driving force is such that the two phases try to separate ($P_e$ 
always negative). For dihedral angles less than $60^\circ$ there exists a 
critical porosity at which the effective pressure is zero, and hence the 
topology is stable. The smaller the dihedral angle, the larger the 
critical 
porosity. If the partially molten medium has a porosity less than this 
critical porosity there will be a tendency of melt to be drawn in, whereas 
if it is above this critical porosity there will be a tendency for melt to 
be expelled. 

In Park \& Yoon's original study, it was found that solutions with zero 
effective pressure could be found for dihedral angles up to $75^\circ$, 
rather than the $60^\circ$ limit found here. One difference between Park 
\& Yoon's study and the present study is that Park and 
Yoon \cite{Park1985} consider a 
unit cell taking the shape of a rhombic dodecahedron. However, the 
discrepancy with the results of this study is likely to have arisen from 
Park \& Yoon's approximation of the grain-melt surfaces by circular arcs, 
which 
is not consistent with the surfaces being of constant mean curvature.

\section{Geometrical properties}

\begin{figure}
 \centering
 \includegraphics[width=\columnwidth]{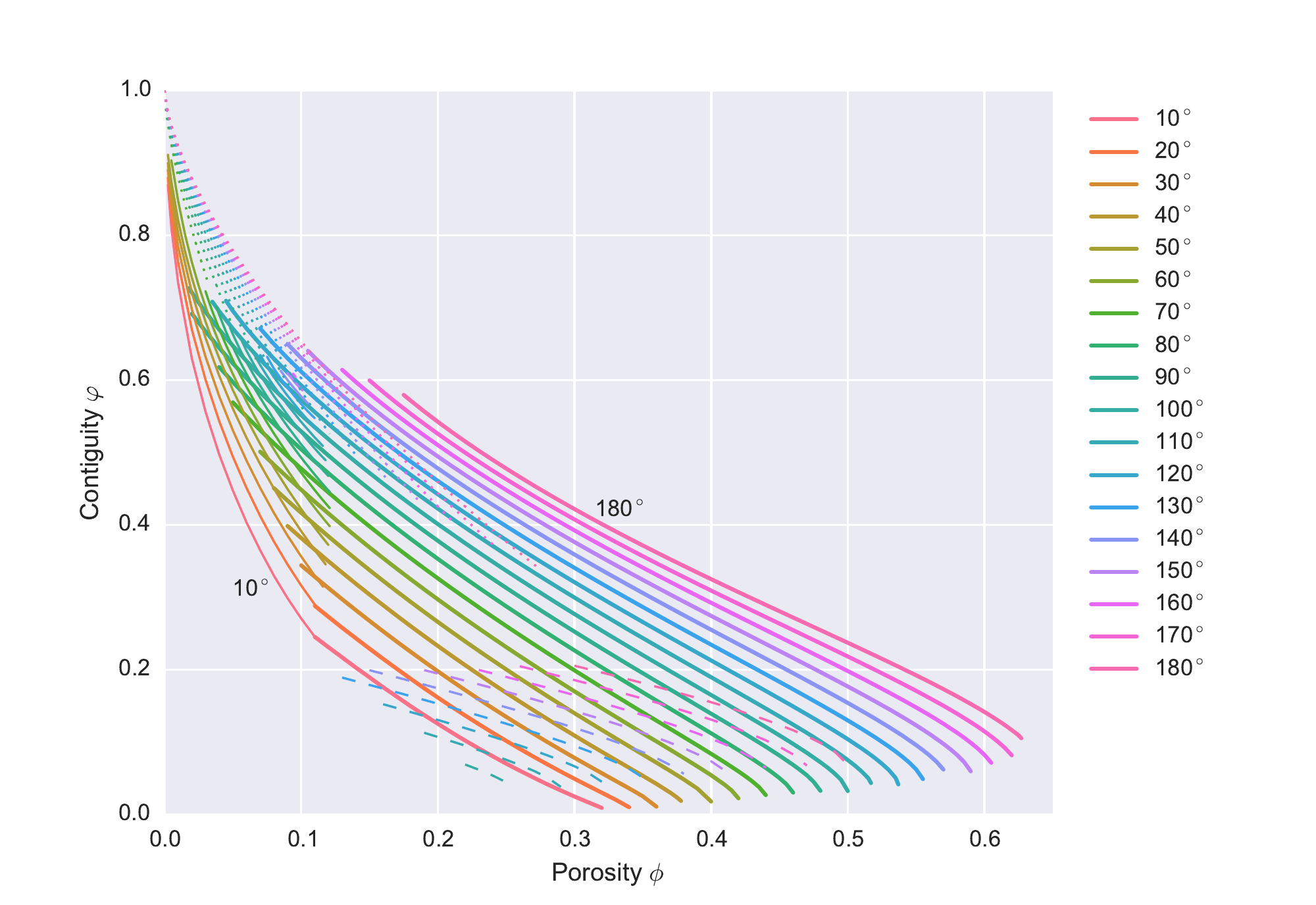}
 \caption{Contiguity. $\varphi = A_\text{ss} / \left(A_\text{ss} + 
A_\text{sl} 
\right)$ }
 \label{fig:contiguity}
\end{figure}

\begin{figure}
 \centering
 \includegraphics[width=\columnwidth]{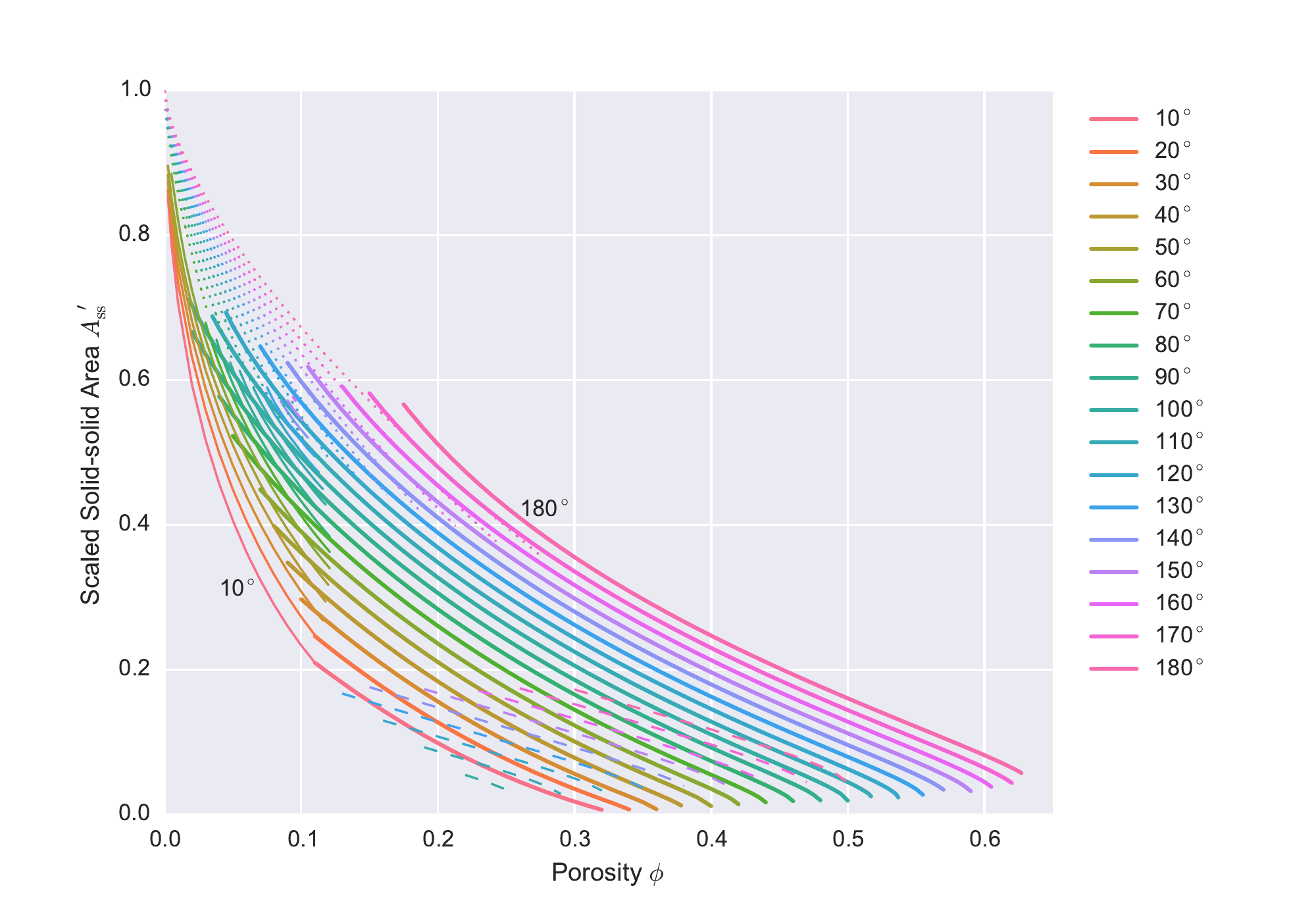}
 \caption{Total area of solid-solid contact, normalised to area of unit 
cell, 
$A_\text{ss}/A_\text{cell}$.}
 \label{fig:solidsolid}
\end{figure}

\begin{figure}
 \centering
 \includegraphics[width=\columnwidth]{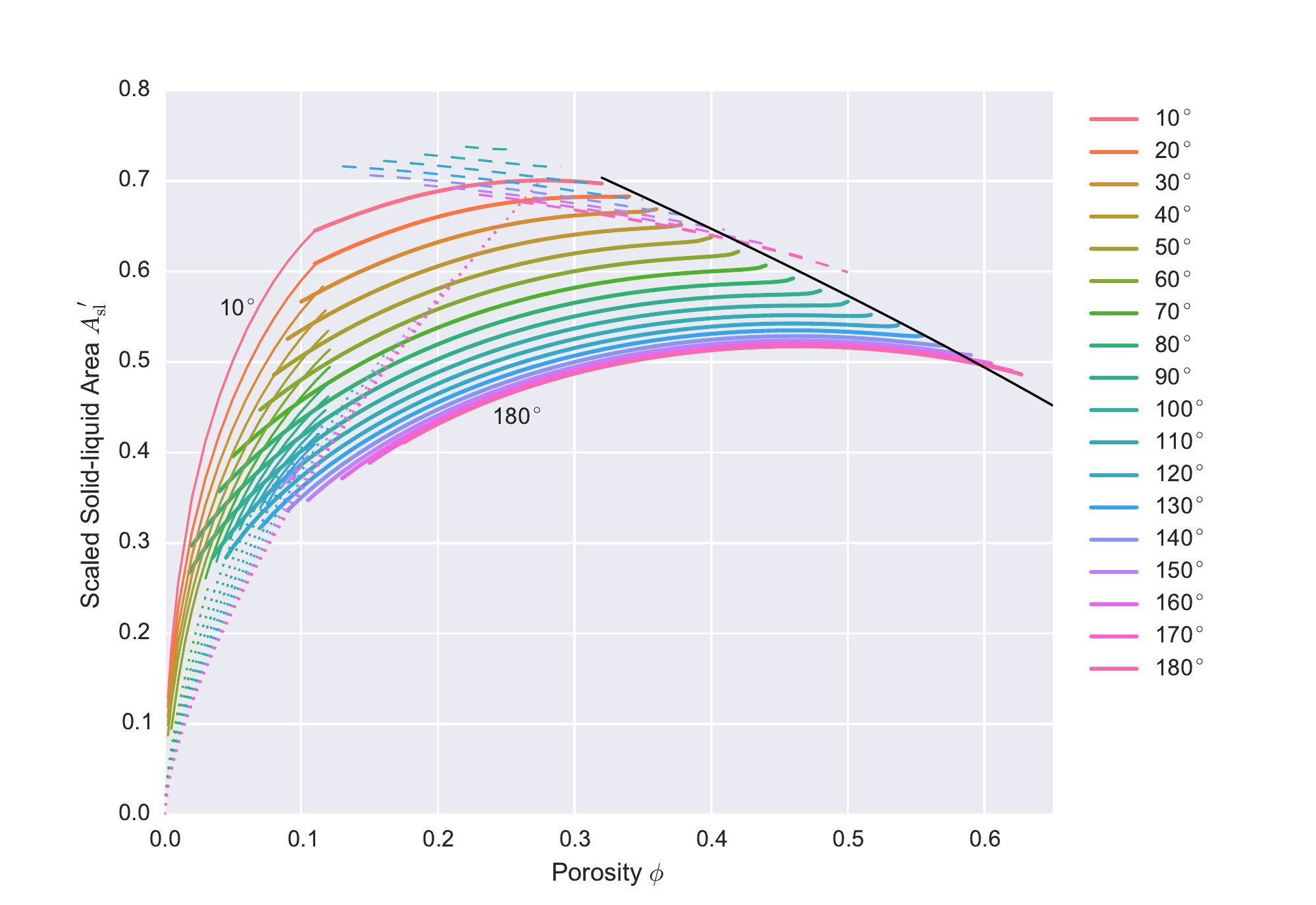}
 \caption{Total area of solid-liquid contact, normalised to area of unit 
cell, 
$A_\text{sl}/A_\text{cell}$.}
 \label{fig:solidfluid}
\end{figure}

\begin{figure}
 \centering
 \includegraphics[width=\columnwidth]{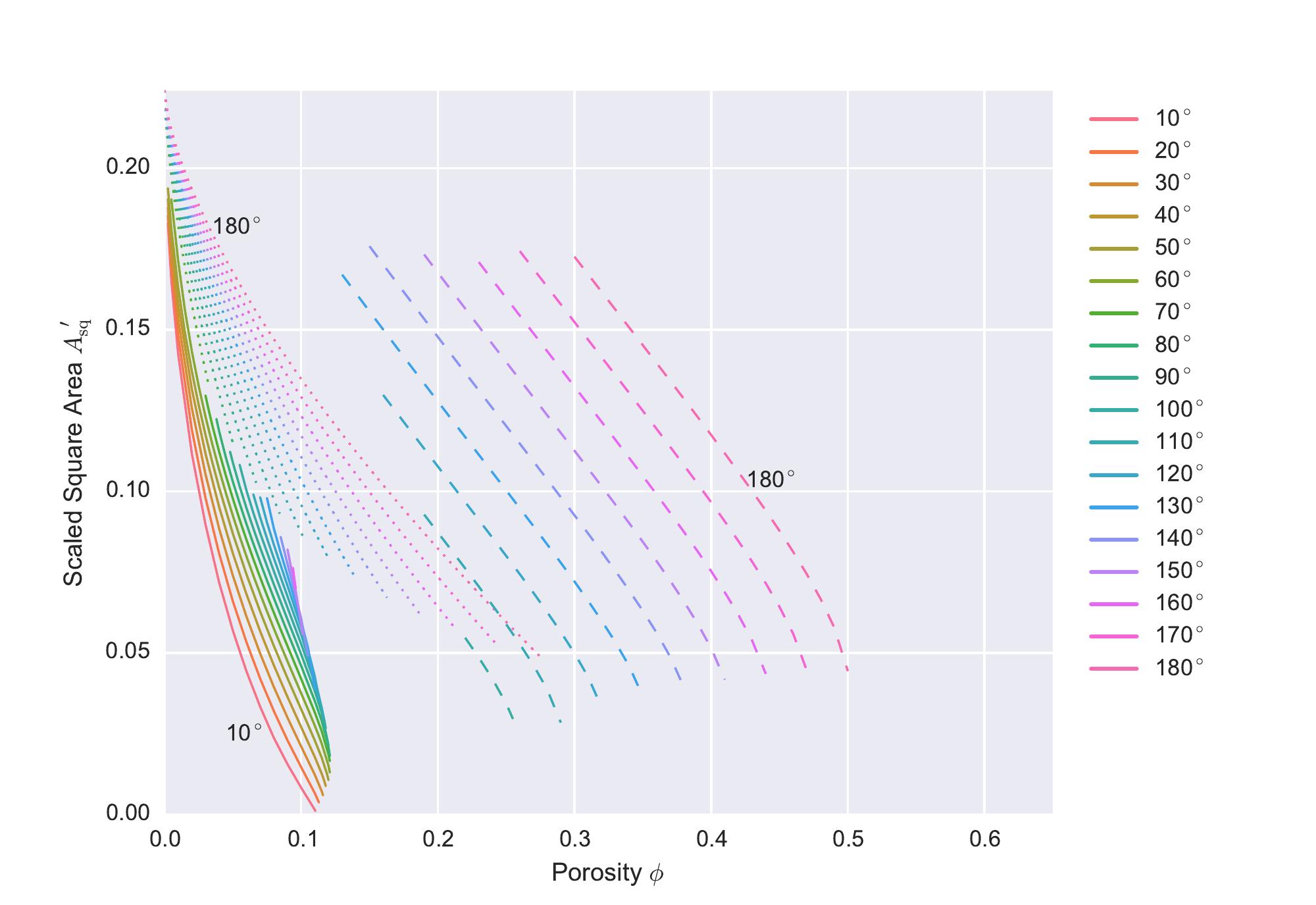}
 \caption{Square solid-solid contact area, normalised to area of unit 
cell, 
$A_\text{sq}/A_\text{cell}$. The ``s'' and ``d'' topologies have zero 
square contact area and 
are not shown.}
 \label{fig:square}
\end{figure}

\begin{figure}
 \centering
 \includegraphics[width=\columnwidth]{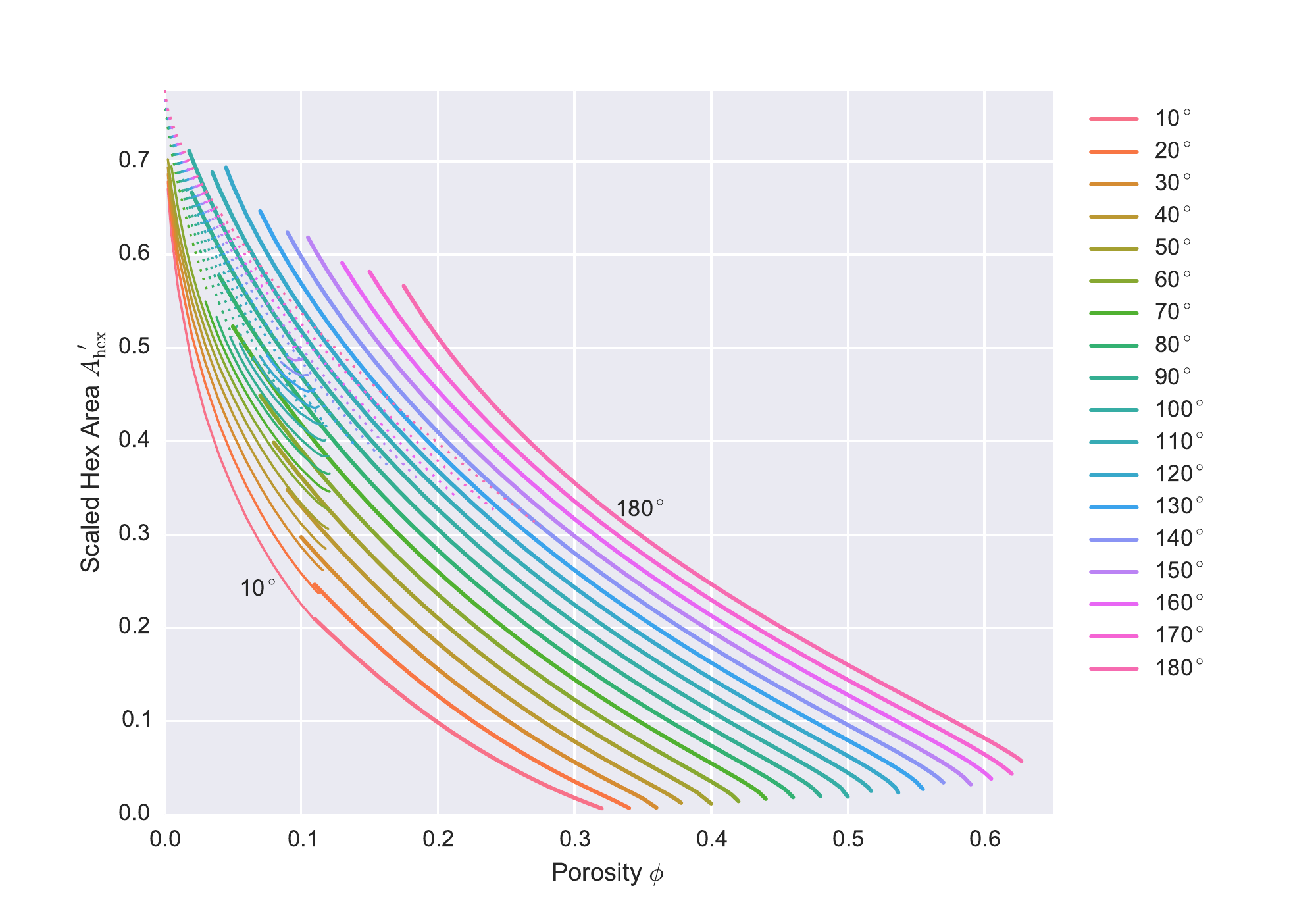}
 \caption{Hexagonal solid-solid contact area, normalised to area of unit 
cell, 
$A_\text{hex}/A_\text{cell}$. The ``h'' and ``d'' topologies have zero 
square contact area 
and 
are not shown.}
 \label{fig:hex}
\end{figure}

\begin{figure}
 \centering
 \includegraphics[width=\columnwidth]{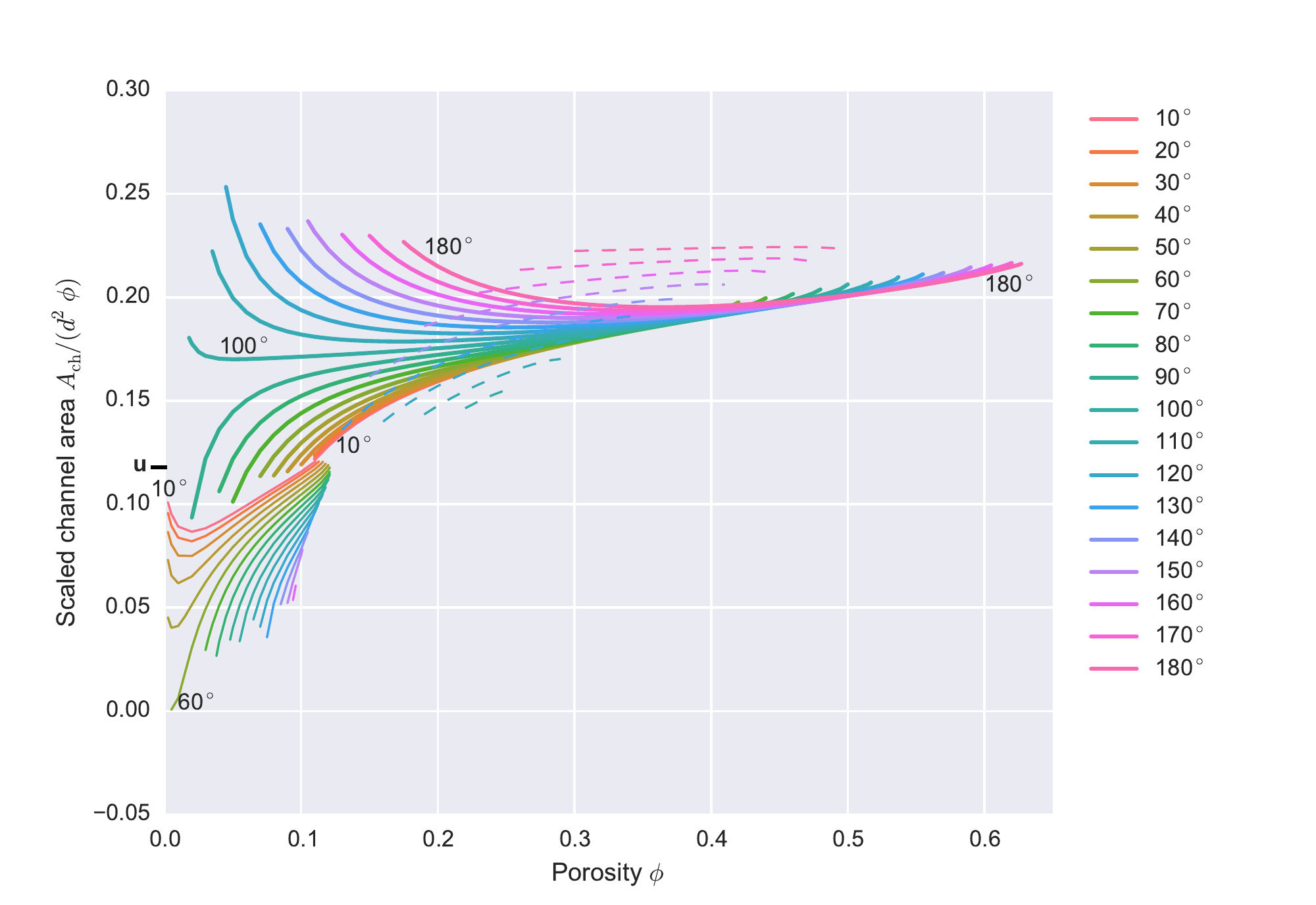}
 \caption{Scaled cross-sectional channel area (sectioned at mid-grain 
edge), scaled as 
$A_\text{ch}/(d^2 \phi)$. Isolated topologies ``i'' have no melt at 
mid-grain edges and thus have zero area and are not shown. Black line 
labelled ``u'' on axis shows the expected channel area for melt tubes of 
uniform 
cross-section along the grain edges.}
 \label{fig:channel}
\end{figure}

Figures \ref{fig:contiguity}-\ref{fig:channel} summarise a number of key 
geometrical properties. The first of 
these (\autoref{fig:contiguity}) shows contiguity $\varphi$,
\begin{equation}
 \varphi = \frac{A_\text{ss}}{A_\text{ss} + A_\text{sl} 
}
\end{equation}
 which measures the relative area of grain-grain contact to total grain 
area. Figures \ref{fig:solidsolid} and \ref{fig:solidfluid} show the 
individual areas of solid-solid and solid-liquid contact, normalised to 
the area of the unit cell. 
Contiguity is used as a fundamental variable in a number of 
micromechanical 
models, e.g. in determining elastic properties \cite{Takei1998} and 
effective 
viscosities due to creep \cite{Takei2009}. For 
small porosities, \autoref{fig:contiguity} shows the same main trends as 
observed by von Bargen and Waff \cite{vonBargen1986} and Cheadle
\cite{Cheadle1989}, with contiguity 
being larger for larger dihedral angles. Contiguity shows a steady 
decrease 
with increasing porosity as more of the grain becomes wetted. There is a 
notable change in slope during the transition from ``c'' to ``s'' 
topologies, 
with a shallower slope for the ``s'' topologies than the ``c'' topologies. 

In this problem there are two different kinds of grain-grain contact -- 
those 
associated with the square faces and those associated with the hexagonal 
faces. 
\autoref{fig:square} plots the areas of the square faces, and 
\autoref{fig:hex} 
the hexagonal faces. For the ``c'' topologies, the areas of the square 
faces 
decreases as porosity increases, but note that they do not vanish at the 
boundary between the ``c+s'' and ``s'' topologies: thus wetting of the 
square 
faces occurs discontinuously from a finite area of square grain-grain 
contact to a zero area as porosity increases. Only for zero dihedral angle 
does 
the ``c'' topology smoothly go to zero area of the square face as porosity 
increases. Such discontinuous jumps in behaviour are common in area 
minimisation 
problems. The classic example of this is in the minimal surface of 
revolution 
problem: finding the soap film which minimises the area between two 
parallel 
circular hoops \cite{Sagan1992}. Beyond a critical separation, the 
catenoid solution which 
joins 
the two hoops breaks down to the Goldschmidt discontinuous solution with 
two 
separate films spanning the two circles.  For small dihedral angles, there 
is a close agreement between these results and those for wet Kelvin foams: 
Hexagonal and square areas reported in Figure 5 of \cite{Murtagh2015} are 
very similar to what one would expect on extrapolating the results of 
Figures \ref{fig:square} and \ref{fig:hex} to zero dihedral angle. 

\autoref{fig:channel} plots a scaled version of the channel area 
$A_\text{ch}$ along the middle of each channel edge (i.e. the area of the 
truncated faces of the quadruple junctions shown in Figures 
\ref{fig:connected} to \ref{fig:wet_hex}). If the melt resided in tubes 
along the grain edges of uniform cross-section, one would expect 
$A_\text{ch} \approx \sqrt{2}/12 d^2 \phi \approx 0. 117851 d^2 \phi$ for 
small porosities (black line on axis in \autoref{fig:channel}). For small 
porosities, and small dihedral angles, 
\autoref{fig:channel} shows that this value is approached, reflecting the 
fact that for such porosities and dihedral angles the melt geometry is 
well 
approximated by tubes. At small porosities, the mid-edge channel area 
decreases with increasing dihedral angle reflecting the fact that more 
melt resides near the grain vertices than the grain edges. The opposite 
trend is seen for the ``s'' topologies at large porosity, where the larger 
dihedral angles have larger channel areas.

\section{Permeability}

\begin{figure}
 \centering
 \includegraphics[width=\columnwidth]{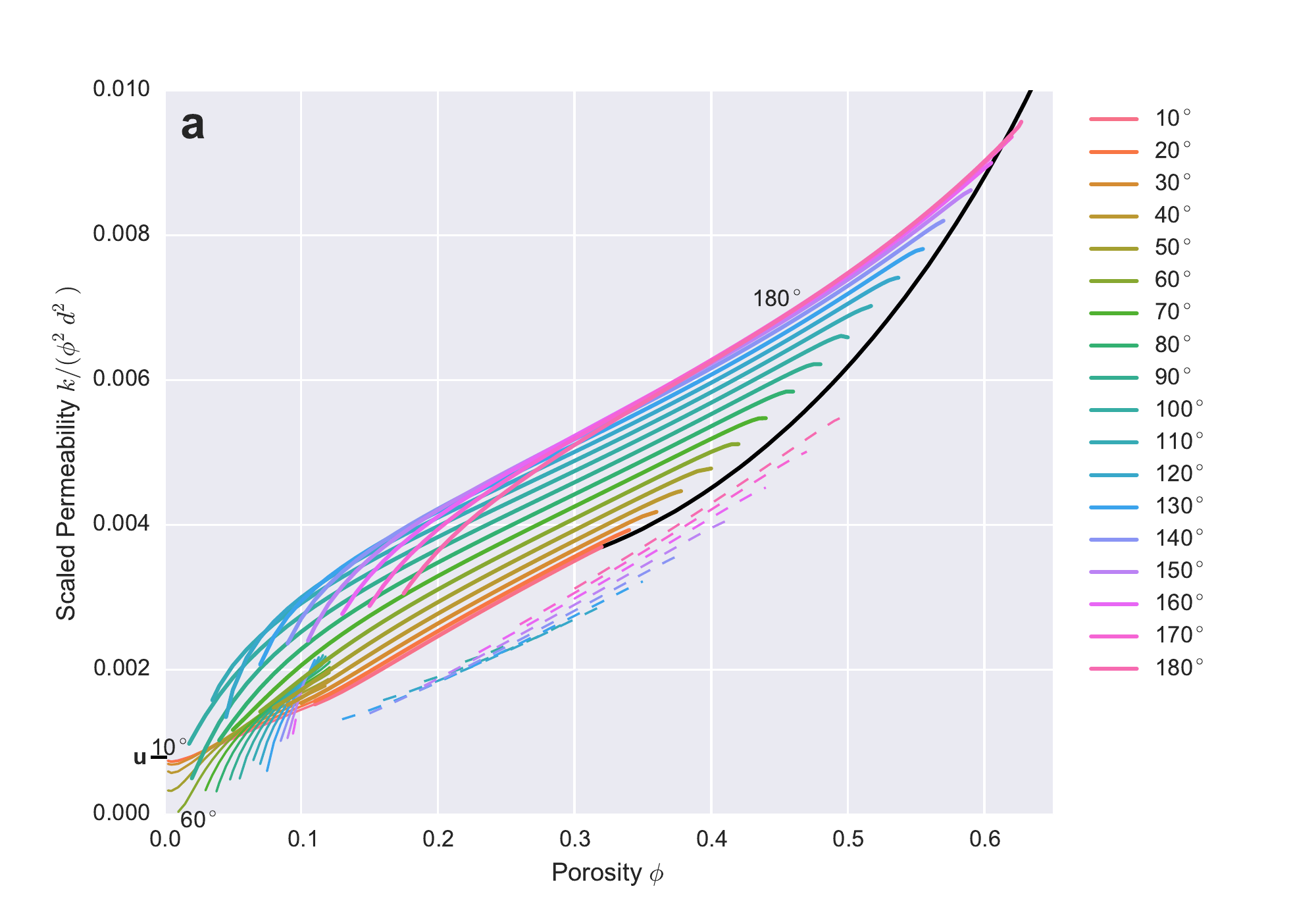}\\
 \includegraphics[width=\columnwidth]{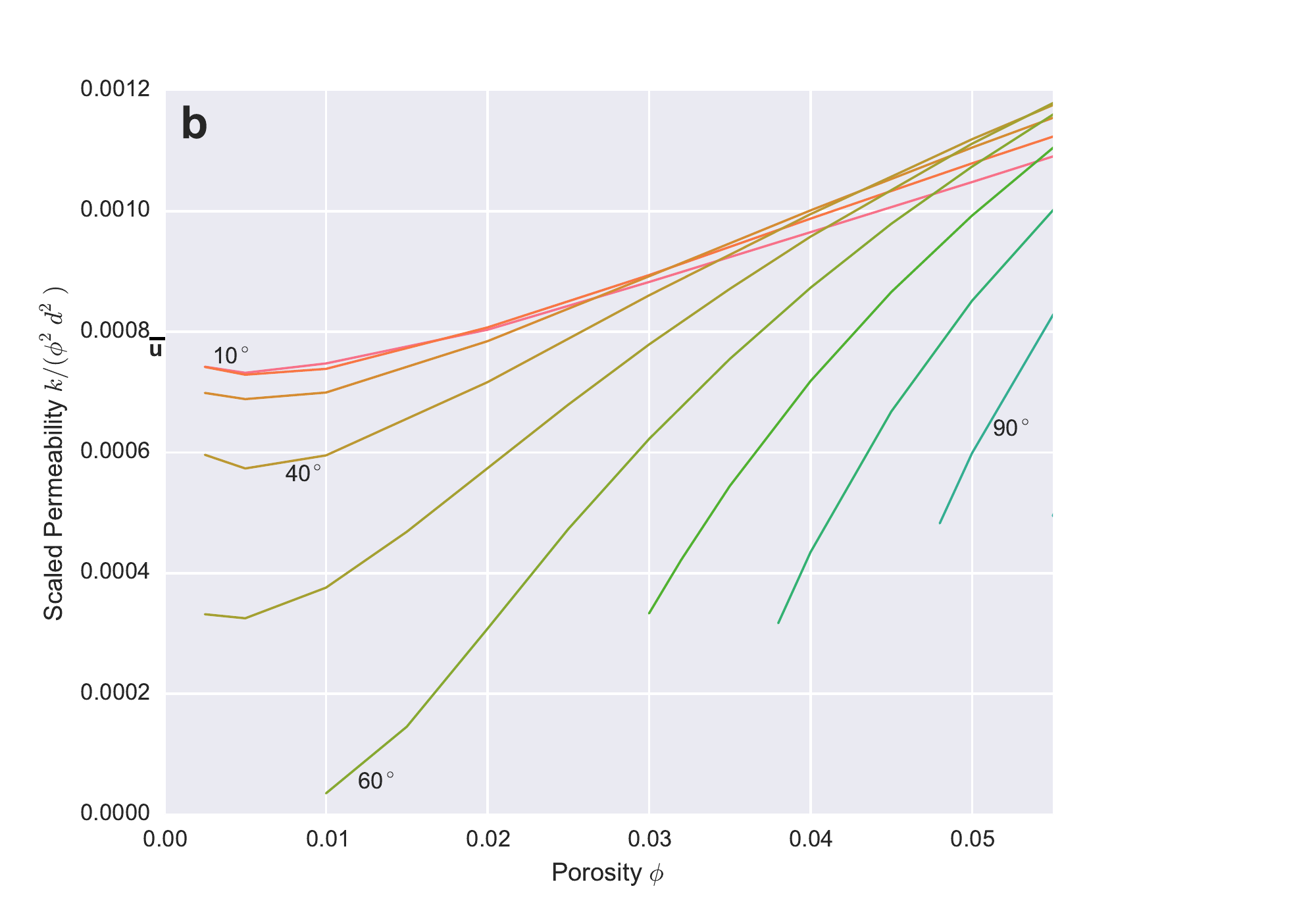}
 \caption{a) Scaled permeability $k / \left(d^2 \phi^2 \right)$. Solid 
black 
line in plot is for topology ``d'', a bcc array of isolated spheres (the 
permeabilities calculated in this case are in agreement with previous 
calculations, e.g. Chapter 9 of \cite{Auriault2009}). Short black line 
labelled ``u'' on axis shows the expected permeability for melt tubes of 
uniform cross-section along the grain edges. b) Same data as shown in 
panel a, but zoomed-in to show the behaviour of ``c'' topologies at low 
porosities (similar to Figure 14 of von Bargen and 
Waff \cite{vonBargen1986}).}
 \label{fig:perm}
\end{figure}

The effective permeabilities of the melt topologies are shown in 
\autoref{fig:perm}, and details of their calculation can be found in 
\autoref{sec:permcalc}. Permeabilities are scaled in \autoref{fig:perm} by 
$d^2 \phi^2$, based on the expected behaviour for low porosities (see 
below).  Most of the main features of \autoref{fig:perm} can be understood 
in the context of simpler models. For small porosities, and small dihedral 
angles, the melt geometry for topology ``c'' takes the form of tubes along 
the grain edges. A very simple model of permeability was derived by 
Frank \cite{Frank1968} consisting of tubes of melt with circular 
cross-section 
lying on the edges of a tetrakaidecahedron. For this model, the 
permeability is
\begin{equation}
 k_\text{Frank} = \frac{\phi^2 d^2}{144 \pi \sqrt{2}} \approx 0.00156 
\phi^2 d^2.
\end{equation}
As discussed by von Bargen and Waff \cite{vonBargen1986}, this formula 
gives an overestimate 
of the permeability at low porosity for the melt networks given here, 
because the melt tubes do not have circular cross-section. For a 
cross-sectional shape more appropriate for low dihedral angles (three 
circular arcs meeting at zero dihedral angle), the melt flux for a given 
cross-sectional area and pressure gradient is a factor of 0.505 lower. A 
modified version of Frank's formula is then
\begin{equation}
k_{\text{tubes}, \theta \approx 0^\circ} \approx 0.000789 \phi^2 d^2 = 
\frac{\phi^2 d^2}{1267},
\end{equation}
and this formula is marked as ``u'' on the axis of \autoref{fig:perm}. 
The equation above explains the limiting behaviour seen in 
\autoref{fig:perm} for low 
porosities and low dihedral angles, which tend to finite values of 
$k/(\phi^2 d^2)$ as $\phi \rightarrow 0$, and whose geometries can be well 
described by tubes 
along the grain edges with approximately uniform cross-section. For larger 
dihedral angles the cross-sections become much less uniform, with more 
melt residing at the vertices of grains rather than the edges. The effect 
of this is to reduce the permeability. von Bargen and 
Waff \cite{vonBargen1986} provide an 
approximate formula,
\begin{equation}
k_{\text{vBW}} \approx 0.000625 \phi^2 d^2 = \frac{\phi^2 d^2}{1600} 
\end{equation}
which is in good agreement with the results here for small porosities and 
dihedral angles around 40$^\circ$. Such a formula gives the permeability 
to within 15\% for all porosities less than 2\%. The expression quoted by 
Cheadle \cite{Cheadle1989}  ($k = {\phi^2 d^2}/{3000}$) seems to 
underestimate 
the permeability, and the permeabilities calculated here are much closer 
to 
the results of von Bargen and Waff \cite{vonBargen1986}. 

For larger porosities ($>10\%$), lines on \autoref{fig:perm} are roughly 
linear, which reflects the fact at large porosities permeability scales as 
$k \propto d^2 \phi^3$. The $\phi^3$ behaviour can also be understood from 
simpler models, and is the expected behaviour for melt lying as thin 
sheets along the grain boundaries. A reasonable approximation for $0.1 < 
\phi <0.3$ and the ``s'' topologies is 
\begin{equation}
 k_{\text{large } \phi} \approx \frac{\phi^3 d^2}{75}
\end{equation}
which lies within about 25\% of the calculated values for dihedral angles 
$\theta <60^\circ$. Indeed, it should be noted that the relative range in 
permeability is very small when $\theta$ varies for porosities above 10\%: 
a factor of 2 at 
most. It is interesting to note that the behaviour of permeability 
with dihedral angle differs at higher porosities: for high porosity, 
the larger the dihedral angle, the greater the permeability; although the 
magnitude of this effect is rather modest. 

Permeability data are often parameterized with a single law of 
the form $k = \phi^n d^2 / C$ (e.g. \cite{Wark1998,Miller2014}). Note 
that such a parametrisations in terms of a single $C$ and $n$ value would 
not be able to describe all the subtleties seen in \autoref{fig:perm}, and 
particularly the shift from $k \propto d^2 \phi^2$ at low porosities to $k 
\propto d^2 \phi^3$ at larger porosities. 

\section{Discussion}\label{sec:discuss}

As remarked in the introduction, the textural equilibrium geometries 
calculated here have also been calculated recently by 
Ghanbarzadeh et al. 
\cite{Ghanbarzadeh2014,Ghanbarzadeh2015,Ghanbarzadeh2015b,
Ghanbarzadeh2016,Ghanbarzadeh2017}
, although 
these authors consider a more limited range of porosities and dihedral 
angles. Ghanbarzadeh et al. \cite{Ghanbarzadeh2015} introduce a novel 
approach to calculating 
textural equilibrium geometries based on the level-set method. In this 
approach, the interface between the solid and liquid is represented by the 
level set of a function, and the interface evolves at a velocity 
proportional to the surface Laplacian of mean curvature. Over time, the 
surface should approach a state of constant mean curvature. A partial 
differential equation determines the evolution of the level-set function, 
and additional terms are added to the PDE to enforce the dihedral angle 
constraint. Ghanbarzadeh et al. \cite{Ghanbarzadeh2015} discretise the PDE 
using high-order 
finite differences.

One of the advantages of Ghanbarzadeh's approach is that it allows a great 
deal of geometric flexibility, including the ability to solve for many 
grains at once with different grain shapes. As such, they do not 
explicitly impose symmetry on the solution (as is done here). Their method 
is considerably more computationally expensive than the approach taken 
here (which can solve for the melt geometries in seconds). 

In broad terms, many of the topologies calculated by 
Ghanbarzadeh et al. \cite{Ghanbarzadeh2015} are similar to those 
calculated here, but there 
are some notable discrepancies. One of the most important discrepancies 
concerns the wetting of the square faces. 
Ghanbarzadeh et al. \cite{Ghanbarzadeh2014} state 
that ``In contrast to prevailing assumptions, the smaller square grain 
boundaries become fully wetted at $\theta=10^\circ$ and $\phi \geq 5 
\%$''. As can be seen in \autoref{fig:regime}, this study places the 
transition to wetted square faces at closer to $\phi=11\%$ for 
$\theta=10^\circ$: a factor of two difference. As remarked earlier, 
transitioning at $\phi=11\%$ is 
consistent with results for wet Kelvin foams 
\cite{Phelan1995,Kraynik2003,Murtagh2015}.

Another discrepancy is in the calculated mean curvatures of the melt 
topologies. Hesse et al. \cite{Hesse2016} presented a plot of mean 
curvature against 
porosity for isolated ``i'' and connected ``c'' topologies for a dihedral 
angle $\theta = 90^\circ$. In theory, this plot should be similar to the 
curvature information shown in \autoref{fig:pressure}, but in fact bears 
little resemblance. First, Hesse et al. \cite{Hesse2016} shows ``c'' 
topologies for 
porosities as low as 1.5\%, whereas the suggested lower bound for ``c'' 
topologies here is around 4\% (very similar to the pinch-off boundary in 
Figure 7 
of von Bargen and Waff \cite{vonBargen1986}). Moreover, the behaviour of 
the mean curvature 
with porosity for the isolated ``i'' topologies shown by Hesse et al. 
\cite{Hesse2016} 
is very different to that seen here. The shape of the isolated topologies 
is independent of porosity, apart from a scaling of the coordinates 
\cite{Wray1976}. Thus mean curvature $H$ should be proportional to 
$\phi^{-1/3}$ for the ``i'' topologies, and be singular at zero porosity 
(as indeed seen 
in \autoref{fig:pressure}). The curve shown by Hesse et al. 
\cite{Hesse2016} does not 
show this behaviour, and instead appears to curve the wrong way, and 
seems to approach a finite value for small porosity.    

These discrepancies matter, because they lead to different predictions 
about upscaled physical properties like permeability. Indeed, the 
permeabilities that have been reported by 
Ghanbarzadeh et al. \cite{Ghanbarzadeh2016,Ghanbarzadeh2017} differ 
from those calculated here by a significant margin, in some cases by 
almost an order of magnitude. For example, for a dihedral angle of 
$10^\circ$, Ghanbarzadeh et al. \cite{Ghanbarzadeh2016,Ghanbarzadeh2017} report that their 
permeability data 
can be fitted well by an expression of the form $k = d^2 \phi^{2.6} / 
595.66$. For a porosity of $\phi=0.01$ this implies $k/(d^2 \phi^2) 
\approx 1 \times 10^{-4}$, whereas here a value of $k/(d^2 \phi^2) \approx 
7 \times 10^{-4}$ has been estimated (\autoref{fig:perm}). Thus 
Ghanbarzadeh et al. \cite{Ghanbarzadeh2016,Ghanbarzadeh2017} seem to underestimate the 
permeability by a factor 
of 7 for these parameter values. For a porosity of $\phi=0.1$, 
Ghanbarzadeh's expression implies $k/(d^2 \phi^2) \approx 
4 \times 10^{-4}$, whereas here $k/(d^2 \phi^2) \approx 
1.4 \times 10^{-3}$ has been estimated, more than a factor of 3 greater. 
Perhaps the most 
likely reason for the discrepancies is that the simulations of 
Ghanbarzadeh et al. \cite{Ghanbarzadeh2015} are under-resolved, and 
struggle to accurately 
capture small scale variations in the geometry e.g. with cusps at small 
dihedral angle, or small radii of curvature at low porosities. The 
accuracy of the solutions obtained in this study is discussed in 
\autoref{sec:numerics}.

Another point of difference between this study and Ghanbarzadeh's is 
that Ghanbarzadeh's study essentially aims to produce surfaces of 
constant mean curvature compatible with the dihedral angle constraints. 
Constant mean curvature is a necessary, but not a sufficient, condition 
for minimum energy. Constant mean curvature guarantees that 
the energy is extremised, but does not say whether the surface is a 
minimum, maximum, or a saddle point of the energy. Indeed, it is 
well-known that some solutions of the Euler-Lagrange equations for area 
minimisation problems can be unstable, as in the
minimal surface area of revolution problem mentioned earlier. Hence it is 
possible that the scheme 
proposed by Ghanbarzadeh et al. produces melt topologies that are not 
minimum 
energy, and hence would not be found by the approach taken 
here. 

The existence of multiple melt topologies has important 
consequences for upscaled physical properties (e.g. permeability, 
electrical conductivity, effective viscosities). First, transitions in 
such properties can be discontinuous on varying parameters, as the 
underlying transitions in topology are discontinuous. Moreover, as 
remarked by von Bargen and Waff \cite{vonBargen1986}, and more recently 
by Hesse et al. \cite{Hesse2016,Ghanbarzadeh2017}, 
there is the potential for hysteresis in topology and hence hysteresis in 
upscaled physical properties. For example, suppose the dihedral angle is 
$40^\circ$ and porosity is increased slowly from zero such that textural 
equilibrium is maintained (\autoref{fig:regime}). The topology would 
remain 
as type ``c'' until a porosity of 12\% when a discontinuous transition to 
type ``s'' topology would occur (wetting of the square faces), when the 
``c'' state is no longer stable. If porosity 
were then slowly reduced, the topology would stay as type ``s'' until a 
porosity of 9\% when a discontinuous transition back to ``c'' would occur, 
as the grain-melt surfaces begin to touch on the square faces (de-wetting 
of the square faces). Similar hysteresis can occur in other parts of the 
regime diagram, for example the ``i'' to ``c'' transition discussed 
by von Bargen and Waff \cite{vonBargen1986}.

There are several natural ways in which the present work can be extended. 
In the same way that this study presents a model with a lower degree of 
symmetry than the studies by von Bargen and Waff \cite{vonBargen1986}, 
Cheadle \cite{Cheadle1989}, 
and Nye \cite{Nye1989}; it would be fruitful to examine the consequences 
of 
relaxing more of the symmetry assumptions. Symmetry plays a crucial role 
in constraining the solutions produced here, and topologies which are 
stable under the symmetry constraints imposed here may not be stable if 
some of these symmetry constraints are relaxed. 

In this study a tetrakaidecahedral unit cell was chosen, but it would be straightforward to repeat the same calculations for other choices of unit cell, such as the rhombic dodecahedral unit cell used by Park and Yoon \cite{Park1985}. In a tessellation of tetrakaidecahedrons, at each vertex four edges meet, and each vertex is thus said to have a coordination number of 4. In a tessellation of rhombic dodecahedrons, half of the vertices have a coordination number of 4, and half have a coordination number of 6. X-ray tomographic imaging suggests that the predominant coordination number in olivine-basalt systems is 4 \cite{Skemer2017}, so the tetrakaidecahedral unit cell is the more appropriate geometry for the partially molten rocks of the Earth's mantle.

To simplify the calculations, grain boundaries were forced to be planar, 
and forced to lie on certain symmetry planes. Grain boundaries are in 
general not perfectly planar, and these constraints could be 
relaxed in future work. Indeed it is the curvature of grain boundaries 
that allows for grain growth and coarsening, an effect than is suppressed 
in the present calculations by the assumptions that have been made. 
Another natural extension is to consider the anisotropy in surface 
energy, both between grains, and between the grains and the melt. 
Anisotropy can lead to faceted solid-liquid and solid-solid interfaces 
(e.g. \cite{Voorhees1984}) that will present challenges for numerical 
simulation. Another challenge for the future is to move beyond 
textural equilibrium, to study the interplay between deformation and melt 
network geometry.

\section{Conclusions}

The aim of this study has been to produce a simple reference model which 
describes the geometry of melt within a partially molten material. This 
model, 
based around a tessellation of tetrakaidecahedral unit cells, has an 
advantage 
over most previous models in that the melt geometry is compatible with a 
space-filling tessellation of grains. A lot of the fundamental 
behaviour of the melt topologies calculated here agrees with the classic 
models of von Bargen and Waff \cite{vonBargen1986}, 
Cheadle \cite{Cheadle1989}, and Nye \cite{Nye1989}, 
particularly at small porosities. The main differences occur at higher 
porosities, where topologies that wet the square faces of the 
tetrakaidecahedral grains exist.

These geometries form a starting point for the 
calculation of upscaled physical properties which depend on the geometry 
of 
melt at the grain scale. One example of such a property has been presented 
here, namely the permeability, but the intent is to describe a wider range 
of physical properties in future manuscripts. In particular, calculations 
on the effective creep properties will be presented, building on the 
simpler grain-scale models of Takei and Holtzmann \cite{Takei2009}.

\subsection*{Acknowledgements}
The ideas for this work arose out of the 2016 ``Melt in the Mantle'' 
programme at the Isaac Newton Institute for Mathematical Sciences. I am 
grateful for 
numerous conversations with the participants of that programme, but 
particularly Marc Hesse for introducing to me to the work of 
Ghanbarzadeh et al. \cite{Ghanbarzadeh2014}, Yasuko Takei for introducing 
me to the work of von Bargen and Waff
\cite{vonBargen1986}, and Dan McKenzie for introducing me to the work of 
Cheadle
\cite{Cheadle1989}. I am also grateful to David Rees Jones for several 
useful discussions. I thank Tom Blenkinsop and an anonymous reviewer for their comments that have helped to improve the manuscript.

\subsubsection*{Funding}
``Melt in the Mantle'' was supported by EPSRC Grant Number 
EP/K032208/1. I am also very grateful to the Leverhulme Trust for 
support.

\subsubsection*{Data Access}Surface Evolver scripts for generating the melt topologies are
available online at \url{www.johnrudge.com/melt}, along with data tables 
of energy, pressure, contact areas, and permeabilities.

\subsubsection*{Competing Interests}
I have no competing interests.

\appendix

\section{Numerical methods}\label{sec:numerics}

\subsection{Problem definition}

The variational problem of minimising $J$ in \eqref{eq:J} is discretised 
by representing the solid-liquid interface by a triangulation as shown in 
\autoref{fig:comp_domain}. Due to the imposed symmetry, points on the 
boundary of the computational domain are constrained to lie on certain 
symmetry planes. In \autoref{fig:comp_domain}a there are five such planes. 
Two of these planes are associated with the triple lines where two 
solid-liquid surfaces meet a solid-solid surface, one corresponding to a 
square face (left), the 
other to a hexagonal face (right). Two more of the symmetry planes are 
where the melt junction is truncated in the figure, marking the middle of 
one of the grain edges. The final plane is an additional mirror plane (top 
left). Of the five symmetry planes, four are mirror planes, but one is 
not: the plane associated with the hexagonal solid-solid contacts. The 
melt surface on the opposite side of the hexagonal triple line is related 
by a $180^\circ$ rotation, not a reflection. This is in contrast to the 
problem of a 
melt junction with tetrahedral symmetry, where the fundamental 
computational domain is bounded solely by mirror planes. Care needs to be 
taken in constructing the mesh so that the edges on the hexagonal triple 
line respect the rotational symmetry. Fortunately, \textit{The Surface 
Evolver} software provides functionality for such symmetry, allowing edge 
elements to ``wrap'' under a symmetry operation (see \cite{Brakke1997} 
for further details).     

All the quantities in $J$ can be calculated by numerical approximations of 
the appropriate integral quantities. The simplest of these is the total 
area of solid-liquid contact, which is the surface integral,
\begin{equation}
 A_\text{sl} = \int_{S_\text{sl}} \d S,
\end{equation}
and when discretised simply involves summing the areas of the individual 
triangles. The solid-solid surface is not triangulated, but its area can 
be calculated from knowledge of its bounding curve (the triple line) 
because of the assumption that solid-solid surfaces are flat. The area is 
given by an application of Stokes' theorem as
\begin{equation}
 A_\text{ss} = \frac{1}{2} \int_{\Gamma_\text{t}} \mathbf{n} \times \left( 
\mathbf{x} - \mathbf{p} \right) \cdot \d \mathbf{x} \label{eq:ss_line}
\end{equation}
where $\mathbf{p}$ is the position vector of the point at the centre of a 
given solid-solid contact, $\mathbf{x}$ is the position vector of a point 
on the triple line, $\mathbf{n}$ is a vector normal to the solid-solid 
interface, and $\Gamma_\text{t}$ is the triple line curve. The volume of 
liquid can be written as a surface integral using the divergence theorem, 
\begin{equation}
 V_\text{l} =  \frac{1}{3} \int_S \mathbf{x} \cdot \mathbf{n} \; \d S,
\end{equation}
where the closed surface $S$ encloses the liquid, and $\mathbf{n}$ is the 
outward normal from the liquid. Part of the surface $S$ is the 
solid-liquid interface, the rest consists of parts of the planes which 
bound the computational domain due to the symmetry. Since these are 
planes, 
 $\mathbf{x} \cdot \mathbf{n}$ is constant over these parts of the surface 
$S$, and the surface integral can be further reduced to a line integral 
similar to \eqref{eq:ss_line}. Furthermore coordinates are chosen so that 
the centre of the melt junction is the origin of the coordinates. Since 
three of the planes (all but the two associated with the truncation) go 
through the origin, $\mathbf{x} \cdot \mathbf{n} = 0$, and there is no 
contribution to the surface integral from these planes. 

\subsection{Optimization}

The Surface Evolver script that solves for the equilibrium geometries 
proceeds by a series of optimization and refinement steps. The initial 
mesh is very coarse, consisting of four triangles  that represent a simple 
planar surface with vertices lying on the five bounding planes. The mesh 
is then alternately optimized, refined, and ``groomed''. Initial stages of 
optimization are performed using gradient decent, later stages are 
performed using Newton's method. The mesh is continually refined until 
individual mesh edges are 0.025 times the grain edge length. ``Grooming'' 
swaps mesh edges so that triangles are roughly equiangular. The 
eigenvalues of the Hessian matrix are checked to ensure all solutions have 
converged to a minimum (with all eigenvalues positive), not a saddle 
point, of the energy. 

\subsection{Different topologies}

The different topologies explored in this contribution have different 
problem setups, essentially due to the removal of one or more of the 
bounding planes from the connected ``c'' topology problem. For example, to 
find ``s''
solutions that wet the square faces, the bounding plane corresponding to 
the square triple line was removed. To find isolated ``i'' solutions, the 
two 
planes associated with truncation were removed. To find hex-wetted ``h'' 
solutions, the plane associated with the hexagon triple line was removed.

\subsection{Accuracy of solutions}

The accuracy of the numerical solutions has been checked by comparing them 
against analytical solutions for combinations of parameters for which such 
solutions exist (see \autoref{app:analytical}). For example, for an ``s'' 
type solution with a dihedral angle of $60^\circ$ that consists of 
spherical surfaces, the numerical mean curvature differs from the 
analytical expression \eqref{eq:ancurve} by only 0.004\%. Since the 
spacing between mesh edges has been kept fixed, the solutions at low 
porosity are liable to be less accurate that those at higher porosity as 
fewer mesh nodes are used. However, the good comparison with the expected 
behaviour at low porosity suggests they are well resolved.  

\subsection{Regime diagram}

To generate the regime diagram depicted in \autoref{fig:regime} a series 
of simulations were run for different parameter combinations in intervals 
of $1\%$ for porosity and $5^\circ$ for dihedral angle. The boundaries 
between different topologies were determined by two different methods 
depending on the nature of the boundary. The first type of boundary occurs 
when the topology assumed in the calculation is inconsistent with the 
tessellation. For example, when calculating the isolated topologies there 
is a point at which the individual isolated pockets of melt meet as 
porosity increases. This can be determined by finding the porosity at 
which 
the extent of the isolated pocket is such that it extends to the middle of 
a grain edge. Similarly, the lower boundary of the ``s'' topology can be 
determined by when the grain-melt surface near the square face first 
touches the boundary of the unit cell. This type of boundary is 
straightforward to determine accurately, and can be obtained through a 
simple interpolation of 
results at different porosities.

\begin{figure}
 \includegraphics[width=\columnwidth]{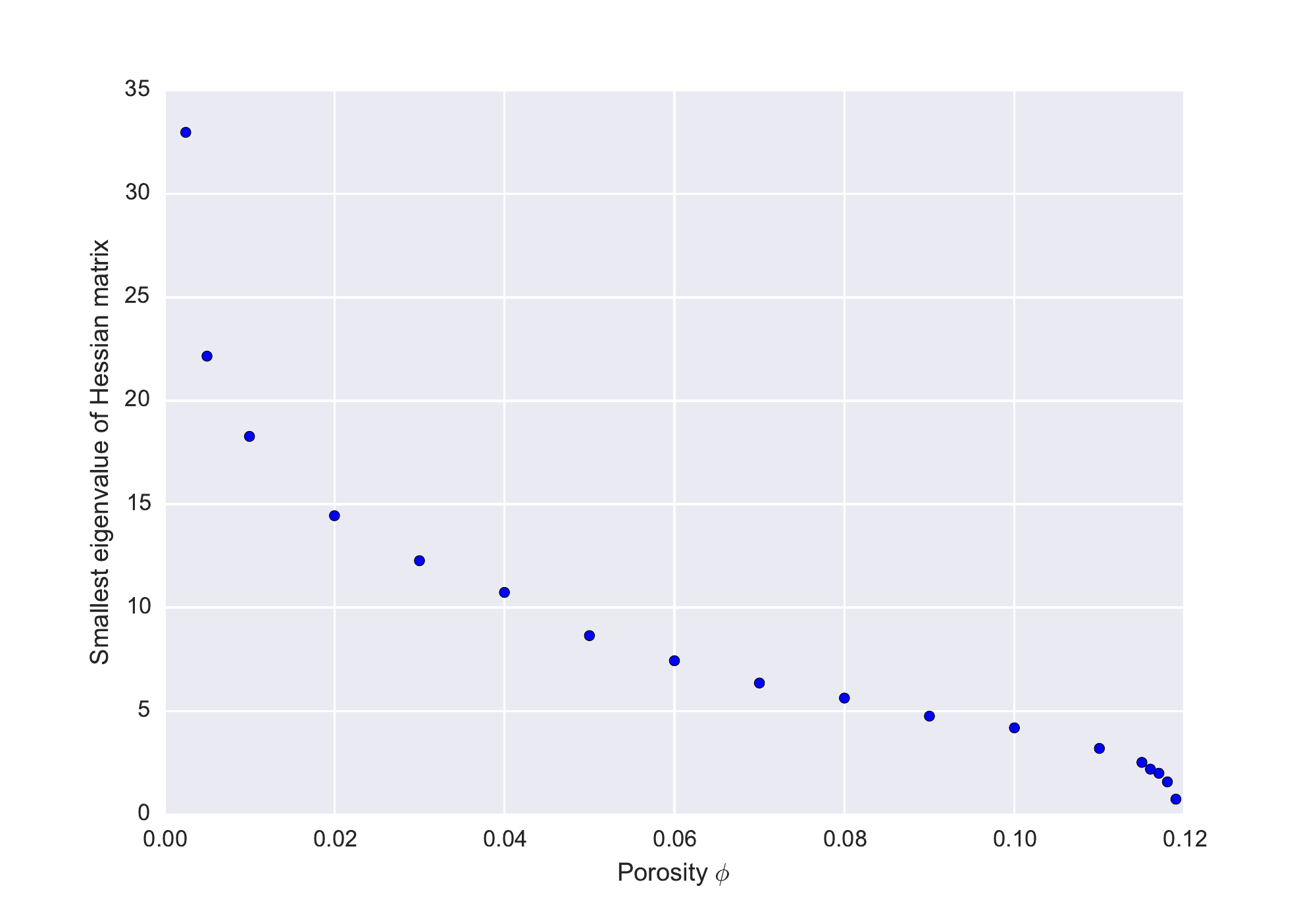}
\caption{Smallest eigenvalue of the Hessian matrix for ``c'' topologies 
and $\theta=40^\circ$. The topology becomes unstable as the porosity 
approaches 12\%.}
\label{fig:evalue}
\end{figure}

The second type of boundary is more difficult to constrain. This occurs 
when the topology in question becomes unstable as the parameters change. 
An example of this type of boundary is where ``c'' topologies no longer 
exist. This type of boundary can be identified by the lowest eigenvalue of 
the Hessian matrix \cite{Brakke1996}, as exemplified by 
\autoref{fig:evalue}. As porosity increases, the lowest eigenvalue 
decreases until some point it reaches zero and the surface is no longer a 
minimum of the energy, but a saddle point. Following \cite{Brakke1996}, 
the boundaries of such topologies have been estimated based on linearly 
extrapolating to the point at which the lowest eigenvalue reaches zero. 
Since the method is based on extrapolation, it is inherently somewhat 
inaccurate. Additional simulations have been run in the vicinity of 
boundaries to improve the accuracy (see additional points near 12\% in 
\autoref{fig:evalue}). Such boundaries are expected to be accurate to 
within at least $\pm0.5\%$ and $\pm 5^\circ$.

\section{Analytical and approximate solutions} \label{app:analytical}

\subsection{Spherical surfaces}

In certain special cases the melt geometry can be calculated analytically, 
and these cases form useful tests of the numerics. These analytical cases 
arise when the solid-liquid interface is spherical. The first of these 
cases is 
for the disaggregated topologies ``d'', where the grains are isolated 
spheres. The energy and curvature are then given by:
\begin{gather}
 E = \gamma_\text{sl} d^2 \pi^{1/3} \left( 3 \left(1 - \phi \right) 
\right)^{2/3}, \\
H = \frac{1}{d} \left(\frac{8 \pi}{3(1-\phi)} \right)^{1/3}. 
\label{eq:sphere_curvature}
\end{gather}
The second case arises for the ``i'' topologies, when the dihedral angle 
is $\theta = 180^\circ$ and the melt forms spherical bubbles at the grain 
corners. Quantities of interest in this case are: 
\begin{gather}
A_\text{sl} = 24 \pi d^2 \left( \frac{\phi}{16 \pi} \right)^{2/3}, \\
A_\text{ss} = \frac{3}{4} \left(1 + 2 \sqrt{3} \right) d^2 - 22 \pi d^2 \left( 
\frac{\phi}{16 \pi} \right)^{2/3}, \\
E = \gamma_\text{sl} A_\text{sl},\\
 H = -\frac{1}{d} \left( \frac{16 \pi}{\phi} \right)^{1/3}.
\end{gather}
Finally, the square-wetted topologies ``s'' are also spherical for 
particular choices of porosity and dihedral angle (see 
\eqref{eq:specialdi} below). The grain takes the 
shape of a sphere with eight spherical caps removed (corresponding to the 
hexagonal faces of the unit cell). Quantities of interest are given by
\begin{gather}
 \phi = 1 + \frac{\sqrt{3} \pi}{16} \left( 6 - 9 \cos \tfrac{\theta}{2} + 
\cos 
\tfrac{3 \theta}{2} \right) \sec^3 \tfrac{\theta}{2}, \label{eq:specialdi} 
\\
H = \frac{4 \cos \tfrac{\theta}{2}}{d \sqrt{3}}, \label{eq:ancurve} \\
A_\text{sl} = \frac{3 \pi}{4} d^2 \sec \tfrac{\theta}{2} \left(4 - 3 \sec 
\tfrac{\theta}{2} \right),\\
A_\text{ss} = \frac{3 \pi}{2} d^2 \tan^2 \tfrac{\theta}{2}.
\end{gather}
These solutions exist from a dihedral angle of zero (where the porosity is 
at the disaggregation value, $\phi = 1- \pi \sqrt{3}/8 \approx 
32\%$) to a dihedral angle of $60^\circ$ where $\phi = 1 + \pi - \pi 3 
\sqrt{3}/4 \approx 6\%$, which lies at the boundary between ``c+s'' and 
``c'' 
in \autoref{fig:regime}.  

\subsection{Tube approximation}

In addition to the analytical solutions provided above, there are some 
approximate solutions that are useful for providing insight into to the 
behaviour at small porosities. One such approximate solution assumes that 
the 
melt resides along tubes of uniform cross section along the grain edges 
(e.g. 
\cite{Frank1968,Koehler2000}). The channel cross-sectional area is 
related to 
the porosity by
\begin{equation}
 A_\text{ch} \approx \sqrt{2}/12 d^2 \phi. \label{eq:tube1}
\end{equation}
For small dihedral angles, a good approximation to the cross-sectional 
shape of 
the channel is given by the region between three touching identical 
circles. In 
this case the radius $r$ of the circles is related to the channel area by
\begin{equation}
 A_\text{ch} = r^2 \left(\sqrt{3} - \frac{\pi}{2} \right). \label{eq:tube2}
\end{equation}
Combining \eqref{eq:tube1} and \eqref{eq:tube2} yields the following 
expression 
for the mean curvature,
\begin{equation}
 H = \frac{1}{d} \sqrt{\frac{3 \left(\sqrt{3} - \pi/2 \right)}{\sqrt{2}}} 
\phi^{-1/2}. \label{eq:tube_curvature}
\end{equation}
Assuming $\gamma_\text{ss} \approx 2 \gamma_\text{sl}$ (appropriate for 
$\theta \approx 0$), other quantities of interest 
can be written as:
\begin{gather}
A_\text{ss} = A_\text{cell} -  3 \sqrt{\frac{2 \sqrt{2}}{ 
\left(\sqrt{3} - \pi/2\right)}}  d^2 \phi^{1/2}, \\
A_\text{sl} = \pi \sqrt{\frac{3 \sqrt{2}}{2 
\left(\sqrt{3} - \pi/2\right)}}  d^2 \phi^{1/2}, \\
 E = \gamma_\text{sl} A_\text{cell} \left( 1- \frac{4}{3} \frac{\sqrt{6 
\sqrt{2} \left(\sqrt{3}-\pi/2 \right)}}{1 + 2 \sqrt{3}}  \phi^{1/2} 
\right).
\end{gather}

\section{Calculation of permeability}\label{sec:permcalc}

The effective permeability of the melt networks was calculated by solving 
the appropriate Stokes flow problem for the unit cell, and then upscaling 
using 
the results from periodic homogenisation theory (e.g. \cite{Hornung1997, 
Auriault2009}). In dimensionless form, the cell problem to solve is
\begin{gather}
\nabla^2 \mathbf{u}_j = \nabla P_j - \mathbf{e}_j, \label{eq:stokes1} \\
\nabla \cdot \mathbf{u}_j = 0 \label{eq:stokes2}
\end{gather}
for velocities $\mathbf{u}_j$, and pressures $P_j$. $\mathbf{e}_j$ is the 
$j^\text{th}$ unit vector ($j = 1, 2, 3$). No-slip ($\mathbf{u}_j = 
\mathbf{0}$) boundary conditions are applied to the 
solid-liquid boundaries, and appropriate periodic boundary conditions are 
applied on the boundaries of the unit cell. The permeability tensor is 
then 
given by
\begin{equation}
 k_{ij} =\frac{1}{V} \int u_{ji} \; \d V,
\end{equation}
where $u_{ji}$ is the $i^\text{th}$ component of velocity $\mathbf{u}_j$, 
and $V$ is the volume of the unit cell. 
Owing 
to the cubic symmetry, the permeability tensor must be isotropic, and 
hence it 
suffices to solve the Stokes flow problem only for $j=1$. A tetrahedral 
mesh of the liquid domain for the unit cell was generated with mesh edge 
size equal to that used for the triangulation of the solid-liquid surface. 
Equations \eqref{eq:stokes1} and \eqref{eq:stokes2} were solved 
numerically using Taylor-Hood finite elements (second order velocity, 
first 
order pressure) with the FEniCS software \cite{Logg2010,Logg2012}.

\bibliographystyle{RS}
\bibliography{newton}

\end{document}